\documentclass[bst/sn-basic]{sn-jnl}

\usepackage{graphicx}%
\usepackage{multirow}%
\usepackage{amsmath,amssymb,amsfonts}%
\usepackage{amsthm}%
\usepackage{mathrsfs}%
\usepackage[title]{appendix}%
\usepackage{xcolor}%
\usepackage{textcomp}%
\usepackage{manyfoot}%
\usepackage{booktabs}%
\usepackage{algorithm}%
\usepackage{algorithmicx}%
\usepackage{algpseudocode}%
\usepackage{listings}%
\usepackage{natbib}
\usepackage{soul}
\usepackage{subcaption}
\usepackage{lineno}

\theoremstyle{thmstyleone}%
\theoremstyle{thmstyletwo}%

\theoremstyle{thmstylethree}%

\begin{document}
\title[Article Title]{Design and preliminary performance study of the broad-band spectrometer detector for POLAR-2}


\author*[1]{\fnm{Jian-Chao} \sur{Sun}}\email{sunjc@ihep.ac.cn}
\author*[1]{\fnm{Jiang} \sur{He}}\email{hejiang@ihep.ac.cn}
\author*[1,2]{\fnm{Shuang-Nan} \sur{Zhang}}\email{zhangsn@ihep.ac.cn}
\author*[1]{\fnm{Shao-Lin} \sur{Xiong}}\email{xiongsl@ihep.ac.cn}

\author[1]{\fnm{Jiang-Tao} \sur{Liu}}\email{jtliu@ihep.ac.cn}
\equalcont{These authors contributed equally to this work.}
\author[1]{\fnm{Yan-Bing} \sur{Xu}}\email{xuyb@ihep.ac.cn}
\equalcont{These authors contributed equally to this work.}
\author[1]{\fnm{Jia} \sur{Ma}}\email{majia@ihep.ac.cn}
\equalcont{These authors contributed equally to this work.}
\author[1]{\fnm{Shuo} \sur{Wang}}\email{shuowang@ihep.ac.cn}
\equalcont{These authors contributed equally to this work.}
\author[3]{\fnm{Lei} \sur{Shuai}}\email{shuail@ihep.ac.cn}
\equalcont{These authors contributed equally to this work.}
\author[3]{\fnm{Xiu-Zuo} \sur{Liang}}\email{liangxz@ihep.ac.cn}
\equalcont{These authors contributed equally to this work.}
\author[4]{\fnm{Hong-Bang} \sur{Liu}}\email{liuhb@gxu.edu.cn}
\equalcont{These authors contributed equally to this work.}
\author[4]{\fnm{Fei} \sur{Xie}}\email{xief@gxu.edu.cn}
\equalcont{These authors contributed equally to this work.}
\author[5]{\fnm{Ming} \sur{Zeng}}\email{zengming@mail.tsinghua.edu.cn}
\equalcont{These authors contributed equally to this work.}
\author[6]{\fnm{Philipp} \sur{Azzarello}}\email{Philipp.Azzarello@unige.ch}
\equalcont{These authors contributed equally to this work.}
\author[7]{\fnm{J\"org} \sur{Bayer}}\email{bayer@astro.uni-tuebingen.de}
\equalcont{These authors contributed equally to this work.}
\author[6]{\fnm{Franck} \sur{Cadoux}}\email{Franck.Cadoux@unige.ch}
\equalcont{These authors contributed equally to this work.}
\author[8]{\fnm{Nicolas} \sur{De Angelis}}\email{nicolas.deangelis@inaf.it}
\equalcont{These authors contributed equally to this work.}
\author[4]{\fnm{Huan-Bo} \sur{Feng}}\email{fenghb@gxu.edu.cn}
\equalcont{These authors contributed equally to this work.}
\author[4]{\fnm{Zu-Ke} \sur{Feng}}\email{fengzk@st.gxu.edu.cn}
\equalcont{These authors contributed equally to this work.}
\author[1]{\fnm{Min} \sur{Gao}}\email{gaom@ihep.ac.cn}
\equalcont{These authors contributed equally to this work.}
\author[9]{\fnm{Ramandeep} \sur{Gill}}\email{rsgill.rg@gmail.com}
\equalcont{These authors contributed equally to this work.}
\author[10,11]{\fnm{Jonathan} \sur{Granot}}\email{granot.j@gmail.com}
\equalcont{These authors contributed equally to this work.}
\author[12]{\fnm{Jochen} \sur{Greiner}}\email{jcg@mpe.mpg.de}
\equalcont{These authors contributed equally to this work.}
\author[7]{\fnm{Alejandro} \sur{Guzman}}\email{guzman@astro.uni-tuebingen.de}
\equalcont{These authors contributed equally to this work.}
\author[13,1]{\fnm{Jin-Xiu} \sur{Hu}}\email{hujx@ihep.ac.cn}
\equalcont{These authors contributed equally to this work.}
\author[1]{\fnm{Yue} \sur{Huang}}\email{huangyue@ihep.ac.cn}
\equalcont{These authors contributed equally to this work.}
\author[6]{\fnm{Johannes} \sur{Hulsman}}\email{Johannes.Hulsman@unige.ch}
\equalcont{These authors contributed equally to this work.}
\author[14,1]{\fnm{Zheng-Huo} \sur{Jiang}}\email{zhjiang@ihep.ac.cn}
\equalcont{These authors contributed equally to this work.}
\author[15]{\fnm{Merlin} \sur{Kole}}\email{merlinkole@gmail.com}
\equalcont{These authors contributed equally to this work.}
\author[3]{\fnm{Dao-Wu} \sur{Li}}\email{lizdao@ihep.ac.cn}
\equalcont{These authors contributed equally to this work.}
\author[16]{\fnm{Han-Cheng} \sur{Li}}\email{hancheng.li@unige.ch}
\equalcont{These authors contributed equally to this work.}
\author[14,1]{\fnm{Tong-Lei} \sur{Liao}}\email{liaotl@ihep.ac.cn}
\equalcont{These authors contributed equally to this work.}
\author[17,1]{\fnm{Long} \sur{Peng}}\email{penglong@ihep.ac.cn}
\equalcont{These authors contributed equally to this work.}
\author[18]{\fnm{Agnieszka} \sur{Pollo}}\email{agnieszka.pollo@ncbj.gov.pl}
\equalcont{These authors contributed equally to this work.}
\author[16]{\fnm{Nicolas} \sur{Produit}}\email{Nicolas.Produit@unige.ch}
\equalcont{These authors contributed equally to this work.}
\author[18]{\fnm{Dominik} \sur{Rybka}}\email{dominik.rybka@ncbj.gov.pl}
\equalcont{These authors contributed equally to this work.}
\author[7,1]{\fnm{Andrea} \sur{Santangelo}}\email{santangelo@astro.uni-tuebingen.de}
\equalcont{These authors contributed equally to this work.}
\author[1,2]{\fnm{Li-Ming} \sur{Song}}\email{songlm@ihep.ac.cn}
\equalcont{These authors contributed equally to this work.}
\author[7]{\fnm{Chris} \sur{Tenzer}}\email{tenzer@astro.uni-tuebingen.de}
\equalcont{These authors contributed equally to this work.}
\author[3]{\fnm{Xiao-Ming} \sur{Wang}}\email{wangxm2@ihep.ac.cn}
\equalcont{These authors contributed equally to this work.}
\author[1]{\fnm{Yuan-Hao} \sur{Wang}}\email{wangyuanhao@ihep.ac.cn}
\equalcont{These authors contributed equally to this work.}
\author[1]{\fnm{Bo-Bing} \sur{Wu}}\email{wubb@ihep.ac.cn}
\equalcont{These authors contributed equally to this work.}
\author[13,1]{\fnm{Pei-Lian} \sur{Wu}}\email{wupl@ihep.ac.cn}
\equalcont{These authors contributed equally to this work.}
\author[6]{\fnm{Xin} \sur{Wu}}\email{Xin.Wu@unige.ch}
\equalcont{These authors contributed equally to this work.}
\author[14]{\fnm{Shuo} \sur{Xiao}}\email{xiaoshuo@gznu.edu.cn}
\equalcont{These authors contributed equally to this work.}
\author[1]{\fnm{Sheng} \sur{Yang}}\email{yangsheng@ihep.ac.cn}
\equalcont{These authors contributed equally to this work.}
\author[1]{\fnm{Lai-Yu} \sur{Zhang}}\email{zhangly@ihep.ac.cn}
\equalcont{These authors contributed equally to this work.}
\author[17,1]{\fnm{Lei} \sur{Zhang}}\email{zhanglei@ihep.ac.cn}
\equalcont{These authors contributed equally to this work.}
\author[1]{\fnm{Yong-Jie} \sur{Zhang}}\email{zhangyj@ihep.ac.cn}
\equalcont{These authors contributed equally to this work.}

\affil*[1]{\orgdiv{State Key Laboratory of Particle Astrophysics}, \orgname{Institute of High Energy Physics, Chinese Academy of Sciences}, \orgaddress{\street{Yuquan Road, Shijingshan District}, \city{Beijing}, \postcode{100049}, \country{China}}}
\affil[2]{\orgdiv{University of Chinese Academy of Sciences}, \orgname{Chinese Academy of Sciences}, \orgaddress{\street{Yuquan Road, Shijingshan District}, \city{Beijing}, \postcode{100049}, \country{China}}}
\affil[3]{\orgdiv{Beijing Engineering Research Center of Radiographic Techniques and Equipment}, \orgname{Institute of High Energy Physics, Chinese Academy of Sciences}, \orgaddress{\street{Yuquan Road, Shijingshan District}, \city{Beijing}, \postcode{100049}, \country{China}}}
\affil[4]{\orgdiv{Guangxi Key Laboratory for Relativistic Astrophysics, School of Physical Science and Technology}, \orgname{Guangxi University}, \orgaddress{\street{No.100 Daxue East Road}, \city{Nanning}, \postcode{530004}, \state{Guangxi}, \country{China}}}
\affil[5]{\orgdiv{Department of Engineering Physics}, \orgname{Tsinghua University}, \orgaddress{\street{30 Shuangqing Road, Haidian District}, \city{Beijing}, \postcode{100084}, \country{China}}}
\affil[6]{\orgdiv{DPNC}, \orgname{University of Geneva}, \orgaddress{\street{24 Quai Ernest-Ansermet}, \city{Geneva}, \postcode{CH-1205}, \state{Geneva}, \country{Switzerland}}}
\affil[7]{\orgdiv{Institut f\"{u}r Astronomie und Astrophysik}, \orgname{Eberhard Karls Universit\"{a}t},  \orgaddress{\street{Sand 1}, \city{T\"{u}bingen}, \postcode{72076}, \country{Germany}}}
\affil[8]{\orgdiv{Istituto di Astrofisica e Planetologia Spaziali}, \orgname{National Institute for Astrophysics}, \orgaddress{\street{Via del Fosso del Cavaliere 100}, \city{Roma}, \postcode{00133}, \country{Italy}}}
\affil[9]{\orgdiv{Instituto de Radioastronom\'ia y Astrof\'isica}, \orgname{Universidad Nacional Aut\'onoma de M\'exico}, \orgaddress{\street{Antigua Carretera a P\'atzcuaro $\#$ 8701,  Ex-Hda. San Jos\'e de la Huerta}, \city{Morelia}, \postcode{58089}, \state{Michoac\'an}, \country{M\'exico}}}
\affil[10]{\orgdiv{Astrophysics Research Center of the Open university (ARCO)}, \orgname{The Open University of Israel}, \orgaddress{\street{1 University Road}, \city{Ra'anana}, \postcode{43537}, \country{Israel}}}
\affil[11]{\orgdiv{Department of Natural Sciences}, \orgname{The Open University of Israel}, \orgaddress{\street{1 University Road}, \city{Ra'anana}, \postcode{43537}, \country{Israel}}}
\affil[12]{\orgname{Max-Planck Institute for Extraterrestrial Physics}, \orgaddress{\street{Giessenbachstr. 1}, \city{Garching}, \postcode{85748}, \country{Germany}}}
\affil[13]{\orgdiv{College of Nuclear Technology and Automation Engineering}, \orgname{Chengdu University of Technology}, \orgaddress{\street{No. 1, East Third Road, Erxianqiao, Chenghua District}, \city{Chengdu}, \postcode{610059}, \state{Sichuan}, \country{China}}}
\affil[14]{\orgdiv{School of Physics and Electronic Science}, \orgname{Guizhou Normal University}, \orgaddress{\street{No. 116, Baoshan North Road}, \city{Guiyang}, \postcode{550001}, \state{Guizhou}, \country{China}}}
\affil[15]{\orgdiv{Space Science Center}, \orgname{University of New Hampshire}, \orgaddress{\street{8 College Road}, \city{Durham}, \postcode{03824}, \state{New Hampshire}, \country{USA}}}
\affil[16]{\orgdiv{Geneva Observatory, ISDC}, \orgname{University of Geneva}, \orgaddress{\street{16, Chemin d’Ecogia}, \city{Versoix}, \postcode{CH-1290}, \state{Geneva}, \country{Switzerland}}}
\affil[17]{\orgdiv{Key Laboratory of Stars and Interstellar Medium}, \orgname{Xiangtan University}, \orgaddress{\street{Yuhu District}, \city{Xiangtan}, \postcode{411105}, \state{Hunan}, \country{China}}}
\affil[18]{\orgname{National Centre for Nuclear Research}, \orgaddress{\street{ul. A. Soltana 7}, \city{Otwock}, \postcode{05-400}, \state{Swierk}, \country{Poland}}}


\abstract{

POLAR-2, the successor of the POLAR experiment aboard China's Tiangong-2 space lab, is set to be deployed on the China Space Station. The POLAR-2 mission aims to conducting high-precision polarization measurements of high-energy transients with a primary focus on Gamma-Ray Bursts (GRBs), following POLAR's pioneering accurate polarization measurements of GRB prompt emission. One of the key advancements in POLAR-2 is the inclusion of a dedicated Broad-band Spectrometer Detector (BSD) instrument, designed to provide precise measurements of GRB location and spectral parameters, which are critical inputs for accurate polarization analysis of POLAR-2's dedicated High-energy Polarimetry Detector (HPD), which is made of plastic scintillator bars array. BSD employs a coded-aperture mask imaging technique and pixelated GAGG scintillation crystals, offering a wide half-coded field of view of $\sim132^\circ \times 125^\circ$ and an operational energy range of 10–1000 keV. Simulation results indicate that the instrument can achieve a localization accuracy of approximately $1.5^\circ$ for faint GRBs similar to GRB 170817A, satisfying the core requirements of GRB polarimetry with HPD. BSD also has moderate capability for GRB polarimetry, particularly at several hundred keV energy. This paper outlines the preliminary design of BSD and presents an overall evaluation of its expected scientific performance, based on extensive Monte Carlo simulations and preliminary ground-based calibration tests.}




\maketitle

\section{Introduction}\label{sec:intro}

Gamma-Ray Bursts (GRBs) are among the most extreme and energetic phenomena in the universe, serving as unique laboratories for studying physical processes under conditions unattainable on Earth \citep{Piran-04, Kumar-Zhang-15}. Historically categorized by their duration \citep{Kouveliotou+93}, long GRBs ($\gtrsim 2$ s) are linked to the core-collapse of massive, rapidly rotating stars \citep{MacFadyen_1999, Woosley+Bloom2006}, while short GRBs ($\lesssim 2$ s, with exceptions of Type IL bursts~\citep{wangchenwei2025}) are associated with the mergers of compact binary systems such as neutron star pairs or a neutron star and a black hole \citep{Eichler+89, Nakar07, Berger14}. The landmark detection of GW170817 and its electromagnetic counterpart, GRB 170817A, unequivocally confirmed the latter progenitor scenario and ushered in a new era of multi-messenger astronomy \citep{Abbott17a, Goldstein17, Abbott17b, Abbott17c,Li2018}.

Despite significant progress since their discovery over half a century ago, the fundamental physical mechanisms governing GRBs remain elusive. Key open questions concern the composition (the GRB outflow's degrees of magnetization and pair enrichment), acceleration, collimation, and ultimate energy dissipation mechanisms within the relativistic jets. The geometry of the outflow and the configuration of the magnetic field—whether it is globally ordered or small-scale and tangled and in which patterns—are particularly critical yet poorly constrained \citep{Lyutikov+03, Waxman-03, Granot-03}. Measurements of the linear polarization in the prompt gamma-ray emission provide a powerful, direct observational probe to address these questions. Different theoretical models—invoking synchrotron emission from anisotropic electron distributions, or Compton scattering—predict distinct polarization signatures \citep{Lazzati-06, Toma+09, Covino-Gotz-16, Gill+20}. Thus, precise polarimetry can break degeneracies between models and reveal the underlying physics of the jet and its central engine \citep{Granot-Konigl-03, Beloborodov-11, Lundman+18}.

This compelling scientific motivation has driven numerous polarimetry attempts using various space-borne instruments, including BATSE on CGRO \citep{BATSE}, RHESSI \citep{Lin2003}, IBIS/SPI on INTEGRAL \citep{IBIS, SPI_1999, Kalemci+07}, GAP on IKAROS \citep{Yonetoku2010}, and more recently, CZTI on AstroSat \citep{CZT_1, Chattopadhyay+14}. However, measuring gamma-ray polarization is technically challenging. It primarily relies on Compton scattering within the detector, requiring photons to interact at least twice to reconstruct the scattering angle, which is modulated by the polarization of the incident photons. This, combined with the intrinsically low flux of GRB prompt emission and the instrumental constraints—such as limited pixel size, detector thickness, and overall geometrical configuration—has generally led to polarization measurements of low statistical significance ($\sim 2-3\sigma$) and substantial, often poorly quantified, systematic uncertainties~\citep{McConnell-17}. Early claims of high polarization, such as for GRB 021206 \citep{coburn2003polarization}, were later contested \citep{Rutledge, Wigger}. While subsequent missions like GAP \citep{Yonetoku+11, Yonetoku+12} and CZTI reported several detections, a larger, more robust sample of high-fidelity measurements is highly desirable.

A major step forward was achieved by the POLAR experiment \citep{PRODUIT2018}, deployed on China's Tiangong-2 space lab in 2016. The instrument operated successfully but its campaign was terminated after six months due to an irreparable high-voltage power supply failure. During this period, it detected 55 GRBs and provided polarization measurements for 14 of them, constituting the largest and most statistically significant sample to date \citep{Zhang+19a, Burgess+19, Kole+20, kole2022gamma}. Its results revealed a generally low degree of polarization, suggesting a more complex magnetic field configuration that previously assumed if the radiation mechanism is indeed synchrotron. More details about the theoretical interpretation can be found in~\citep{Zhang+19a, Gill-Kole-Granot-21, Lan_2021}. These findings have fundamentally reshaped the theoretical landscape and underscored the need for next-generation instruments with superior sensitivity.

To address this need, the POLAR-2 collaboration was formed, planning to install a significantly enhanced instrument on the China Space Station (CSS) \citep{Hulsman2025k, polar2overview2025}, currently planned  around 2028. POLAR-2 is designed to achieve a substantial increase in effective area over its predecessor, POLAR. This new design adopts a high-sensitivity SiPM array, which mitigates risks similar to those encountered in POLAR. This design improves the lower energy threshold while eliminating the need for high-voltage power supply modules. This enhancement is energy-dependent, being most pronounced at lower energies and reaching a factor of greater than four at higher energies, which will dramatically improve the sensitivity for polarization measurements. The project employs a multi-instrument approach for comprehensive coverage. More sensitive polarization measurement will be performed by the High-energy Polarimetry Detector (HPD, 30--800 keV) \citep{POLAR-2:2021uea}, complemented by the Spectroscopy and Polarimetry Detector (SPD) which is composed of Broad-band Spectrometer Detector (BSD, 10--1000 keV) and Low-energy Polarimetry Detector (LPD, 2–10 keV) \citep{Feng_2024}. Crucially, precise localization and broadband spectroscopy—essential for accurate polarization analysis—are provided by BSD, which will work in conjunction with LPD,
ensuring that POLAR-2 is fully self-sufficient for the spectro-polarimetric characterization of GRB prompt emissions as well as x-ray flares. The HPD and the SPD (composed of BSD and LPD) are separate payloads. They will be installed in close proximity on CSS with aligned pointing to ensure a common field of view for concurrent observations.

This paper focuses on the design and expected performance of the BSD instrument. BSD utilizes a coded-aperture mask imaging (CAMI) technique to achieve precise GRB localization. We present the scientific requirements (Section~\ref{sec:2}), detailed instrumental design (Section~\ref{sec:3}), comprehensive performance evaluations based on Monte-Carlo simulations (Section~\ref{sec:4}), and results from preliminary ground-based calibration tests with a detector prototype (Section~\ref{sec:5}). Finally, we summarize our conclusions and outline future work (Section~\ref{sec:6}).

\section{Instrument requirements}\label{sec:2}

First, precise localization and spectral parameters of GRBs are essential inputs for polarimetry data analysis. Uncertainties in these parameters can introduce significant systematic errors into the polarization measurements. Polarimeters typically require this information either from their own reconstruction capabilities or from external instruments with simultaneous observations and higher precision. This requirement becomes particularly critical for bright GRBs~\citep{An2023,zhangyanqiu2024}, where systematic uncertainties—rather than photon statistics—dominate the measurement errors. Furthermore, accurate localization is a prerequisite for reliable spectral analysis, as the local instrumental response varies with the photons' incident direction.

However, Compton polarimeters such as POLAR-2/HPD and its predecessor POLAR have limited localization capabilities, with typical precisions of about 3.6$^\circ$~\citep{Kole_2023} and 10$^\circ$~\citep{wang2021localization}, respectively. Therefore, the dedicated POLAR-2/BSD instrument is necessary to provide high-precision localization and spectroscopy. Second, BSD must possess a wide Field-of-View (FoV) to effectively capture high-energy transients such as GRBs, enabling simultaneous observations with the HPD and LPD instruments. This allows for joint data analysis of polarization, spectrum, and location across a larger sample of events. Third, BSD should be capable of being triggered on, localizing, and downlinking GRB alerts in near real-time through the relay satellite system with a typical latency of seconds. Although not a strict requirement for the POLAR-2 experiment, this last functionality would greatly enhance synergistic and follow-up observations in the multi-messenger astronomy era.

Based on these considerations, the following key requirements have been defined for the BSD design:
\begin{itemize}
\item Localization precision: Better than 1$^\circ$ for typical bright GRBs (fluence $>1\times10^{-5}$~erg~$\cdot$~cm$^{-2}$ @10-1000~keV in 1~s, BAND: $\alpha$ = -1, $\beta$ = -2.3, E$_\mathrm{peak}$ = 230~keV), to minimize systematic uncertainties in the polarization analysis for HPD and LPD~\citep{Feng_2025} and to facilitate rapid follow-up observations.
\item Energy range: Broadband spectral coverage from 10 keV to 1000 keV, which encompasses the typical energy range of GRB prompt emission where the spectral peak energy ($E_{\mathrm{peak}}$) is commonly observed.
\item Field of view: The FoV should encompass the primary coverage areas of the HPD and LPD instruments. Given the payload constraints, the BSD is designed with a half-coded FoV of approximately 120$^{\circ}$ × 120$^{\circ}$.
\item Sensitivity: The detection sensitivity should be comparable to or exceed that of current leading GRB monitors, i.e., on the order of $\sim$3~ph~$\cdot$~cm$^{-2}$~$\cdot$~s$^{-1}$@10-1000~keV in 1~s.
\end{itemize}

In its current design, BSD incorporates the capability to trigger and localize GRBs in orbit. GRB alert messages will be transmitted quasi-instantaneously to the ground via the space station's fast downlink channel.

\section{Instrument design}\label{sec:3}

To meet the scientific requirements, several detector designs have been evaluated for BSD, including configurations similar to \textit{Fermi}/GBM~\citep{Meegan_2009} and GECAM~\citep{GECAM2020}, and the CAMI technique employed by \textit{Swift}/BAT~\citep{Gehrels_2004} and SVOM/Eclairs~\citep{godet2014}. The CAMI scheme was ultimately selected for its superior GRB localization accuracy, while satisfying the engineering constraints (e.g., size, weight, power) of a medium external payload on the CSS. The use of a pixelated scintillator array enables broad-band spectral measurements, while also providing inherent polarimetric capability through the detection of Compton scattering patterns between pixels. This results in a compact and efficient spectrometer design for the POLAR-2/BSD payload.

A critical design consideration involves the trade-off between FoV and angular resolution. A large FoV is essential for capturing transient GRBs but, within the fixed payload envelope (600 mm $\times$ 600 mm $\times$ 500 mm), necessitates a shorter mask-to-detector distance, which degrades the intrinsic angular resolution. Since the primary requirement is not the utmost localization precision but a level adequate for polarimetry analysis and capable of providing useful alerts for follow-up observations, a systematic optimization was conducted to balance these competing parameters.

\begin{figure}[h]
\centering
\includegraphics[width=0.9\textwidth]{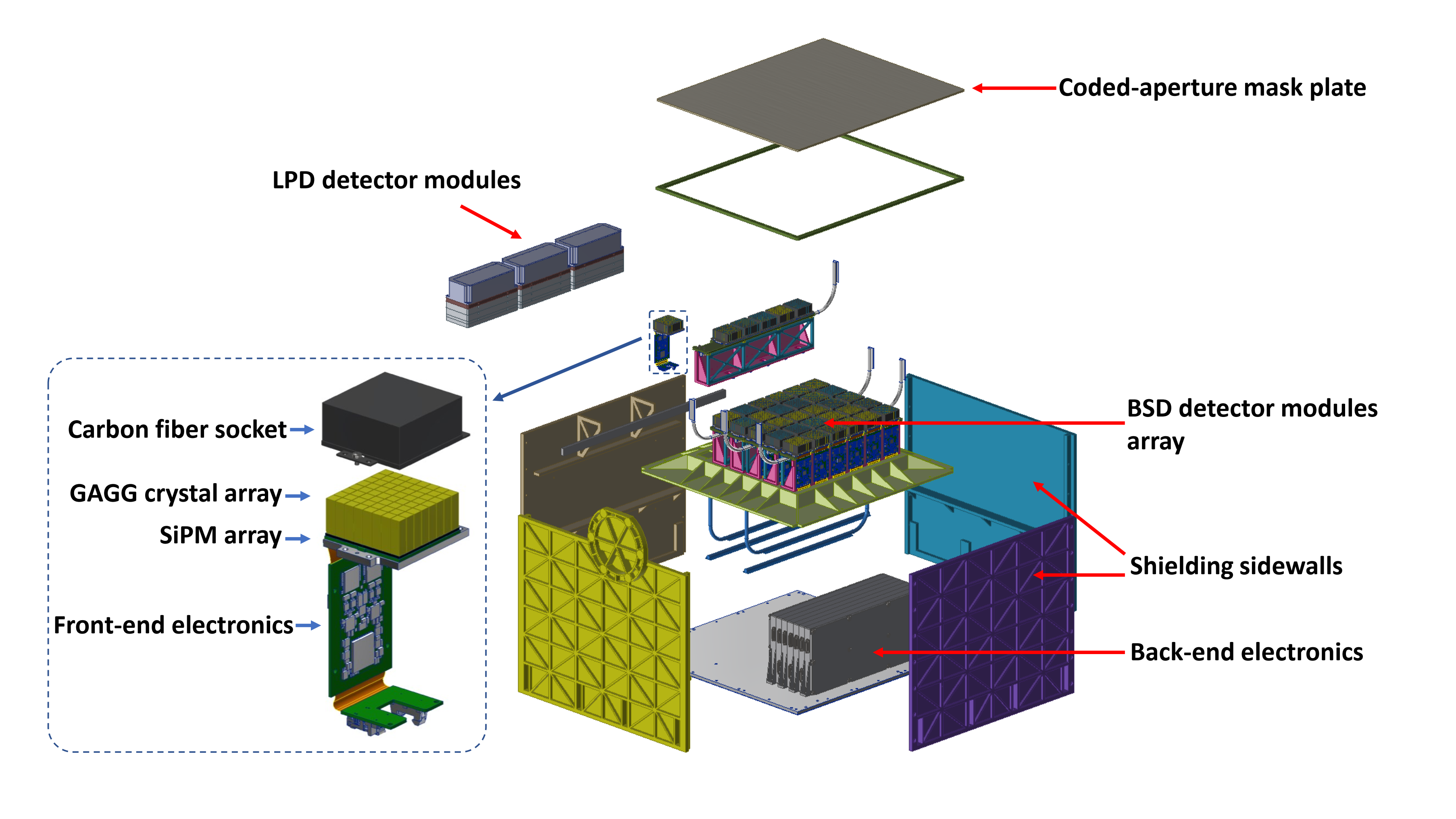}
\caption{Exploded view of the POLAR-2/SPD payload, illustrating the assembly of its key components. A single detector module of BSD is detailed in the dashed box on the left} 
\label{fig:1}
\end{figure}

The final BSD instrument design, illustrated in Fig.~\ref{fig:1}, comprises the following key components:
\begin{itemize}
    \item \textbf{Coded-Aperture Mask Plate}: A mask fabricated from $\sim$3500 tungsten alloy elements (6.25 mm $\times$ 6.25 mm $\times$ 1.00 mm each), arranged in a random pattern with a 50\% open fraction (Fig.~\ref{fig:2}). To enable source localization, the tungsten coded-aperture mask is designed to have energy-dependent transparency, effectively blocking a large fraction of photons below 100 keV while becoming progressively transparent at higher energies due to the penetration of gamma-rays (Fig.~\ref{fig:2-1}). The mask plate is reinforced with low-mass carbon fiber layers ($\sim$1.50 mm thick) to minimize weight. The total mask size is 599 mm $\times$ 499 mm, with an active imaging area of 575 mm $\times$ 475 mm.

\begin{figure}[t!]
\centering
\includegraphics[width=0.7\textwidth]{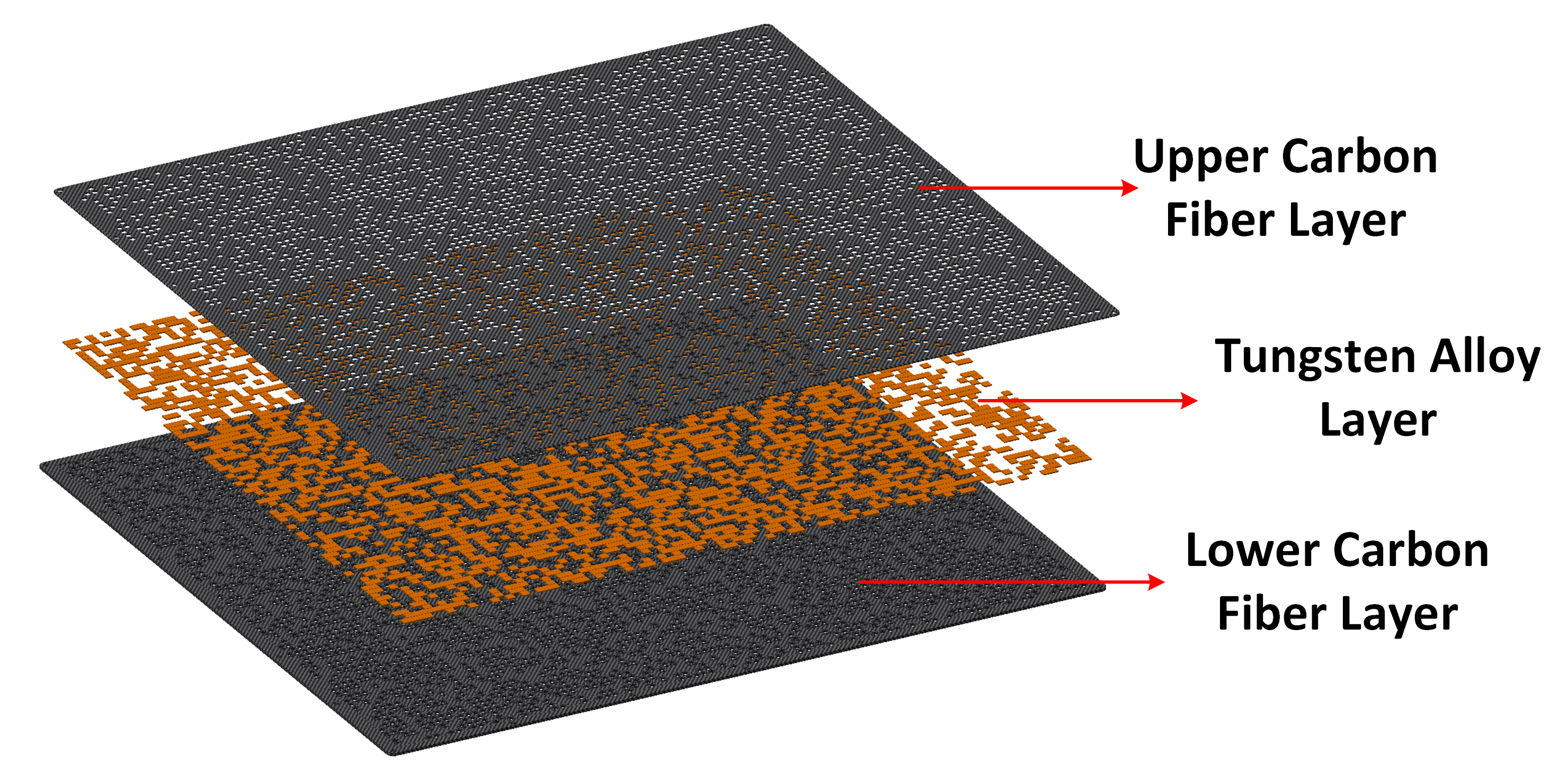}
\caption{Design of the coded-aperture mask plate. The layer of tungsten alloy elements is sandwiched between upper and lower carbon fiber support plates} 
\label{fig:2}
\end{figure}

\begin{figure}[h]
\centering
\includegraphics[width=0.5\textwidth]{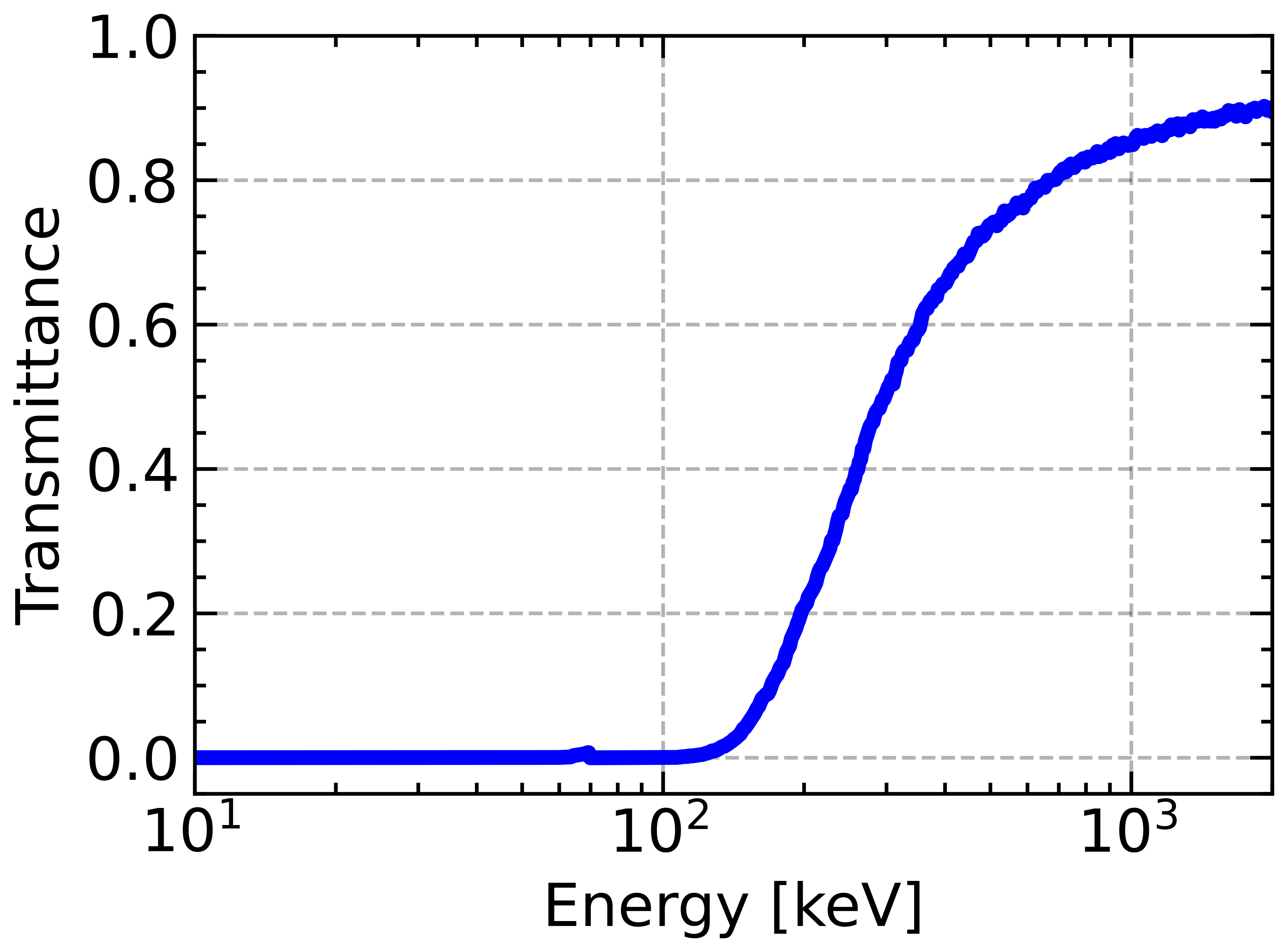}
\caption{Transmittance of the coded-aperture mask plate for photons at different energies, obtained from Geant4 simulations} 
\label{fig:2-1}
\end{figure}
    
    \item \textbf{Detector Array}: The detector plane is located 125.6 mm below the mask and consists of 2304 cerium-doped $\text{Gd}_{3}\text{Al}_{2}\text{Ga}_{3}\text{O}_{12}$ (GAGG:Ce) scintillator crystals (5.75 mm $\times$ 5.75 mm $\times$ 20.00 mm each), providing a total sensitive area of $\sim$761 cm$^2$. The crystals are grouped into 36 modules, each containing 64 pixels in an 8$\times$8 array. The pixel pitch within a module is 6.25 mm, expanding to 18.75 mm between modules—an integer multiple of the fundamental pitch—to accommodate mechanical mounting.
    \item \textbf{Readout System}: Each detector module houses the GAGG:Ce array in a carbon fiber socket. Scintillation light from each crystal is read out by a dedicated Silicon Photomultiplier (SiPM) and front-end electronics (FEE) board, the same design as the HPD payload \citep{kole2025fee} which has been space-qualified by series of environmental tests, such as irradiation, thermal vacuum, vibration and shock tests \citep{DeAngelis2023}.
    \item \textbf{Back-End Electronics (BEE)}: The BEE unit processes data from all 36 modules, hosts the trigger and localization algorithms, generates onboard alerts, and manages telecommands and power distribution. GRB alerts are disseminated to the ground via the CSS's fast downlink channel.
    \item \textbf{Mechanical Structure and Shielding}: The assembly is housed within a structure that provides mechanical support and passive shielding against in-orbit background radiation.
\end{itemize}

\begin{figure}[h]
\centering
\includegraphics[width=0.7\textwidth]{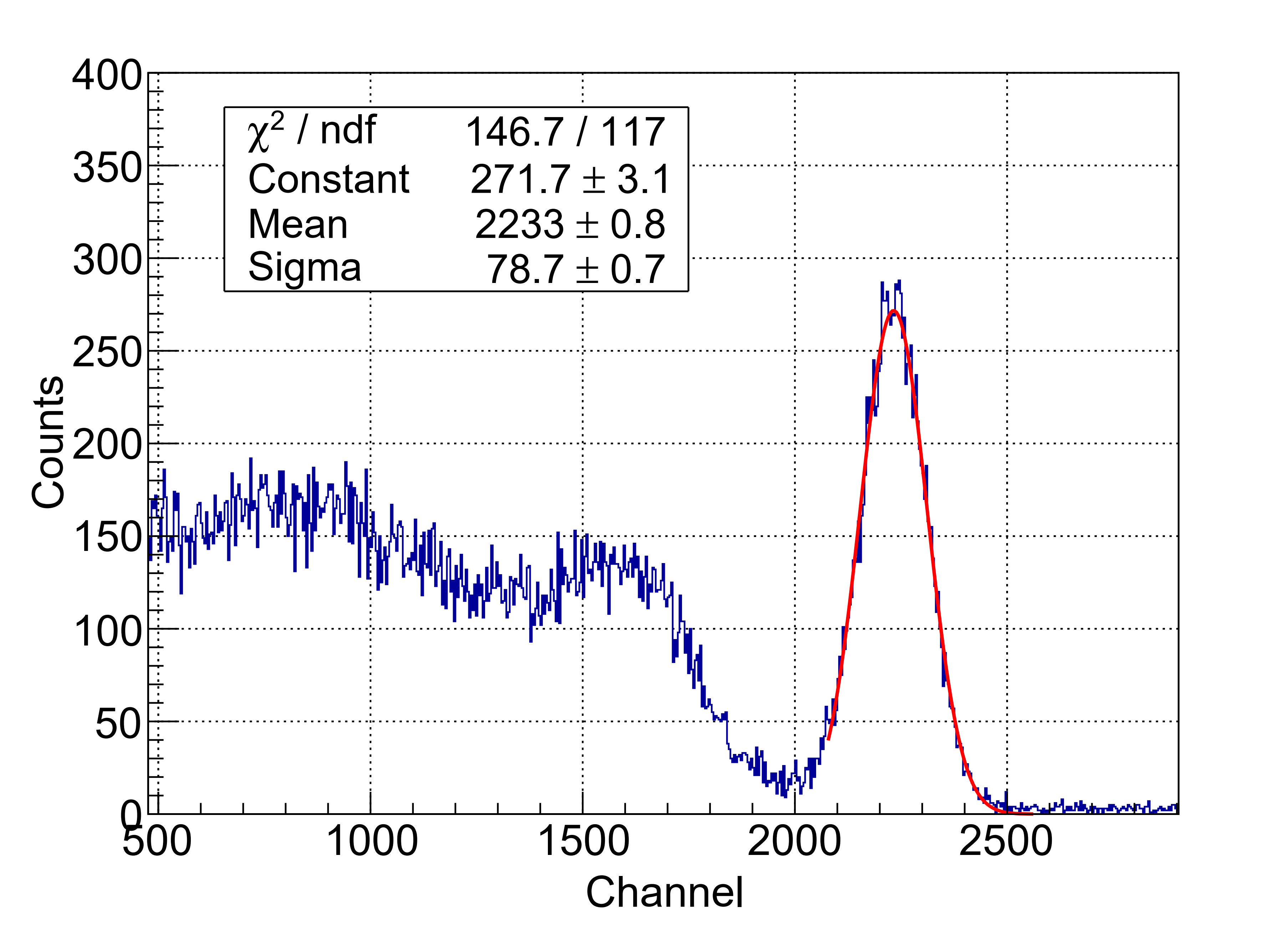}
\caption{Measured energy spectrum (blue curve) of a $^{137}$Cs radioactive source by the BSD detector prototype. By fitting the 662 keV photopeak from the $^{137}$Cs source with the Gaussian function (red line), the energy resolution at 662 keV was determined to be about 8.3$\%$} 
\label{fig:662keV}
\end{figure}

The GAGG:Ce scintillator was selected for its excellent properties, which include a high light yield (often exceeding 50,000 photons/MeV), high density (6.63 g/cm$^3$), fast decay time (less than 100 ns), good energy resolution (a measured value of $\sim$8.3$\%$@662~keV, see Fig.~\ref{fig:662keV}), and non-hygroscopic nature. Its mechanical robustness, lack of intrinsic radioactivity, and ease of handling make it ideal for coupling with SiPMs in pixelated imaging detectors. The crystal's proven performance in space missions such as GRID~\citep{wen2021compact}, and its selection for upcoming missions like SPHiNX~\citep{pearce2019science}, CUBES~\citep{kushwah2021design}, HERMES~\citep{dilillo2024hermes} and CXBe~\citep{Li:20253r}, further validate its suitability for the BSD application. These properties, combined with its high stopping power and cost-effectiveness, facilitate the construction of large-area spectrometers covering a wide energy range (10--1000 keV). The key performance characteristics of the BSD CAMI system are summarized in Table~\ref{tbl:1}.

\begin{table}[h]
\caption{Summary of key technical parameters for the coded-aperture mask imaging system of the POLAR-2/BSD instrument}\label{tbl:1}%
\begin{tabular}{p{2.8cm}p{8cm}}
\toprule
Parameter & Value\\
\midrule
Imaging method      & Coded-aperture mask with random pattern, $\sim$50$\%$ open  \\
Mask cell material  & $\sim$3500 pieces of tungsten alloy, 6.25~mm$\times$6.25~mm$\times$1.00~mm for each piece  \\
Mask frame material & Two layers of $\sim$1.50~mm thick carbon fiber  \\
Mask dimension      & 575~mm$\times$475~mm plus some margin   \\
Lateral shielding material & 0.4~mm thick of tantalum and 1$\sim$10~mm thick of aluminum alloy \\
Detector material   & GAGG:Ce crystal arrays, 5.75~mm$\times$5.75~mm$\times$20.00~mm for each pixel\\
Detector segment    & 36 detector modules, 8$\times$8 GAGG:Ce array for each module  \\
Pitches of detector & 6.25~mm between two adjacent pixels within one module, 18.75~mm between neighboring modules \\
Detecting area      & $\sim$761~$cm^2$   \\
Mask and detector plane distance   & 125.6~mm      \\
Field of view       & $~80^\circ\times~54^\circ$ (full-coded), $~132^\circ\times~125^\circ$ (half-coded)  \\
Average angular resolution & $\sim$1.4$^\circ$ \\
\botrule
\end{tabular}
\end{table}

\section{Performance simulations}\label{sec:4}

Based on the Geant4 simulation toolkit\footnote{\url{https://geant4.web.cern.ch/}}, the performance of the BSD detector was evaluated using Monte-Carlo methods. The software enables detailed mass modeling of the detector structure, as described in Section~\ref{sec:3}, along with the implementation of appropriate physics models to simulate the interaction of incident particles with the instrument.

A comprehensive simulation study was conducted to assess the overall performance of the detector and its compliance with mission requirements. Key performance metrics—including energy response, detection sensitivity, expected GRB detection rate, and localization accuracy—were investigated in accordance with the objectives outlined in Section~\ref{sec:2}. Prior to these analyses, a full mass model of the instrument was constructed, and the in-orbit background environment and corresponding count rates were determined.

\subsection{Mass modeling}\label{subsec:4-1}

The detector simulation, based on the Geant4 (version 10.7.p02) framework, incorporates a dedicated module for constructing the instrument's mass model. This module accurately defines the geometric dimensions and material composition (including atomic constituents and density) of each detector component. For the BSD instrument, the modeling begins with the sensitive detector array, where each unit comprises 64 GAGG:Ce crystals. Each crystal is modeled as a rectangular prism with dimensions of \(5.75 \times 5.75 \times 20 \, \text{mm}\). In the simulations, interaction details of incident particles within the crystals—such as the detector unit, pixel identifier, interaction position, and energy deposition—are recorded for subsequent analysis. For the subsequent simulations, the Geant4 Shielding physics list was utilized, which comprehensively handles electromagnetic and hadronic interactions, including the activation of materials.

\begin{figure}[h]
\centering
\includegraphics[width=0.6\textwidth]{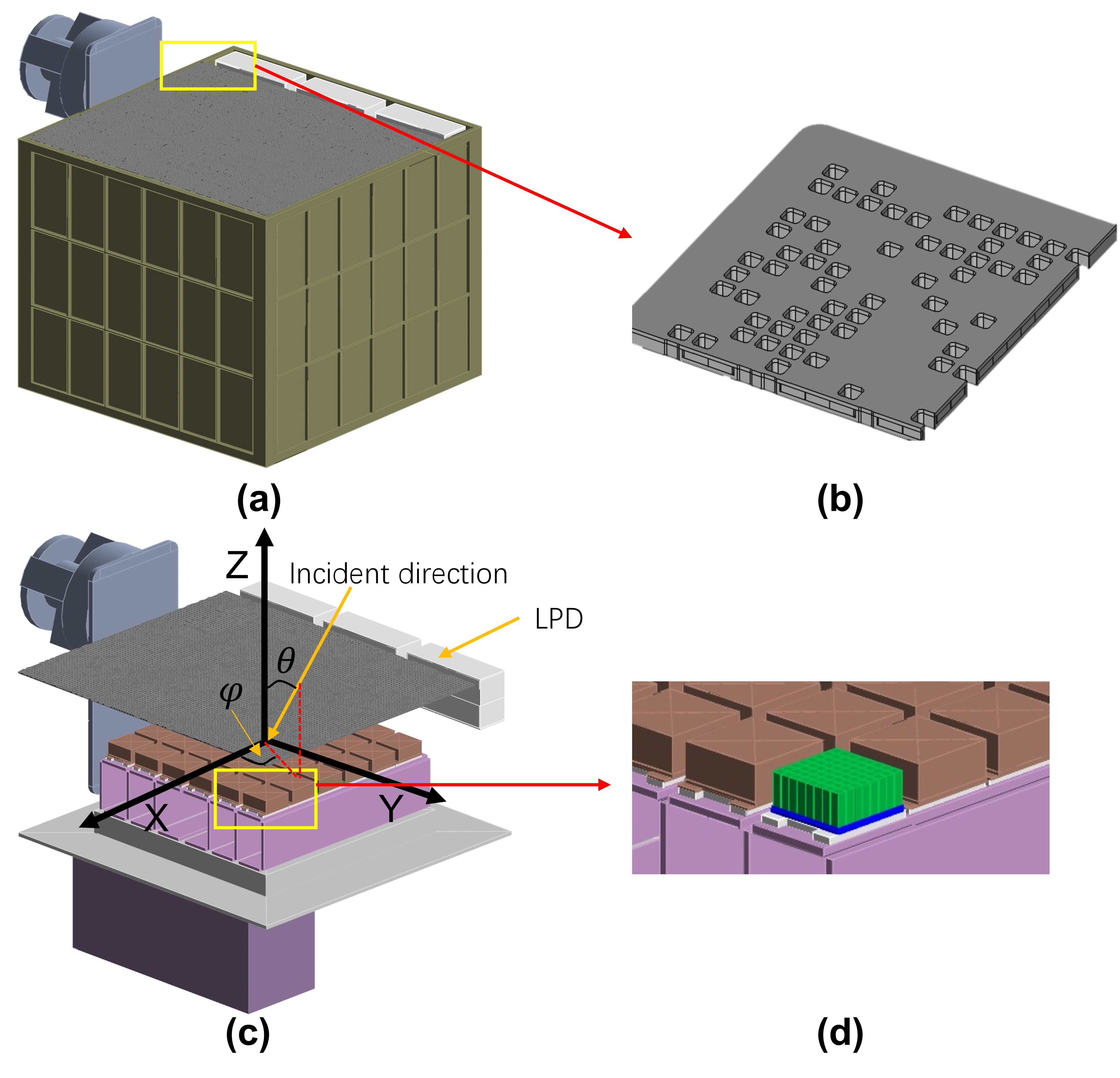}
\caption{Geant4 mass model of the POLAR-2/BSD instrument. (a) Full assembly of the instrument. (b) Close-up view of the coded-aperture mask. (c) Internal structure with shielding walls removed. (d) Detailed view of the GAGG:Ce detector array} 
\label{fig:3}
\end{figure}

The detection process may be influenced by materials surrounding the sensitive volume. Particles may undergo Compton scattering or be fully absorbed within these components before reaching the active detector elements, potentially leading to non-detection. Conversely, these structures also provide shielding against the in-orbit low-energy particle background. Given their substantial impact, a detailed mass model of all external detector components is essential. While justified simplifications can be made based on their expected influence on the detector response, a high-fidelity representation is crucial. The complete Geant4 mass model of the BSD instrument, derived from the professional structural design, is shown in Fig.~\ref{fig:3}.

\begin{figure}[h]
\centering
\includegraphics[width=0.7\textwidth]{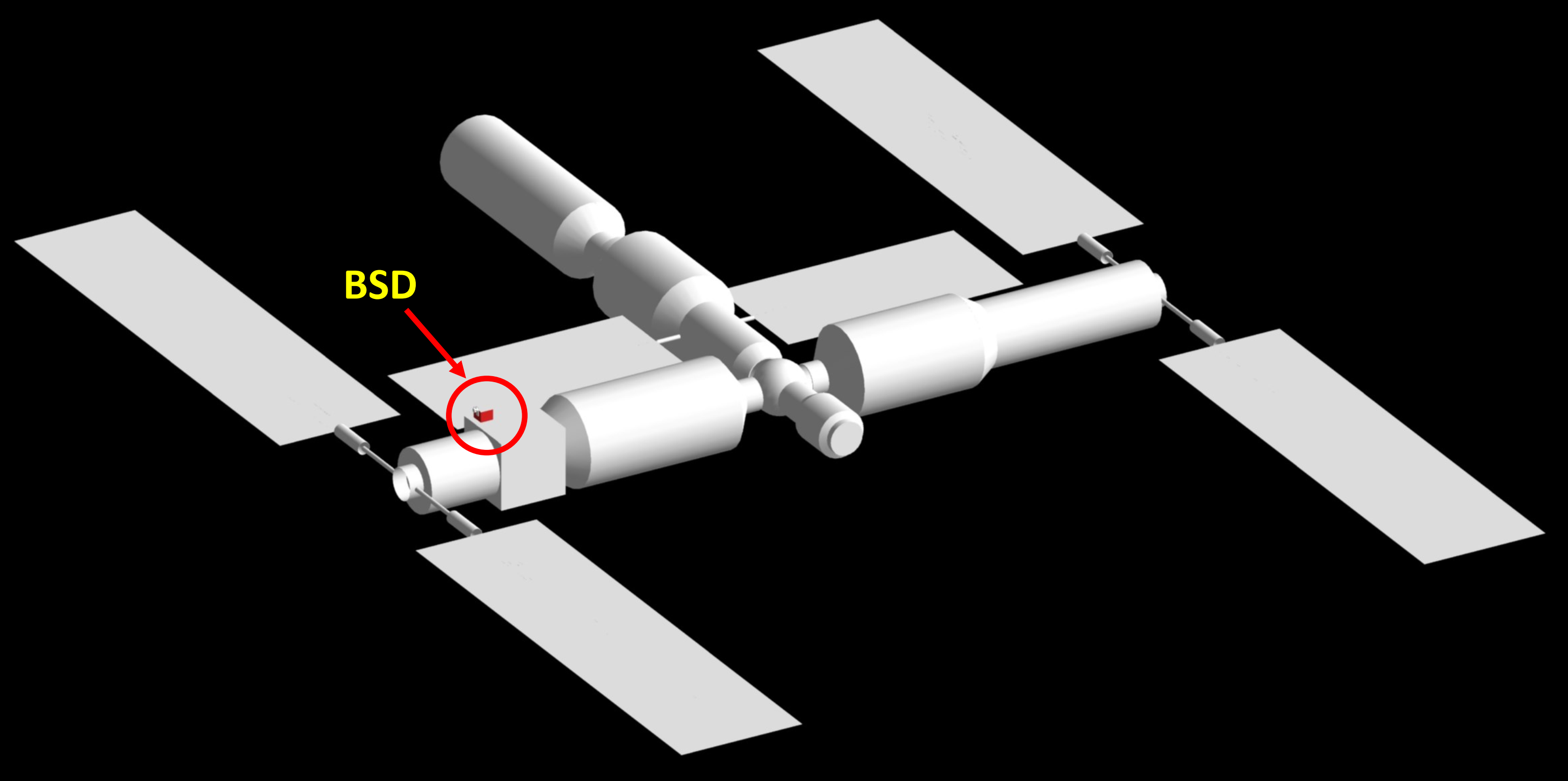}
\caption{Geant4 mass model of the China Space Station (CSS), showing the anticipated installation position of the BSD instrument} 
\label{fig:4}
\end{figure}

To further enhance the simulation's realism, the influence of the CSS platform was included. Structures on the platform can scatter photons into the detector's field of view that would not otherwise hit it directly. A simplified model of the CSS was constructed based on its overall dimensions and material properties. Unknown surrounding payloads were omitted, as this level of simplification is deemed sufficient for the current stage of performance evaluation. The simplified CSS mass model, indicating the anticipated installation position of BSD, is presented in Fig.~\ref{fig:4}.

\subsection{In-orbit background}\label{subsec:4-2}

Maximizing the observational duty cycle is crucial for GRB detectors to capture randomly occurring transient events. Consequently, the dominant source of detected events in orbit is the continuous flux of background particles, which directly impacts key performance metrics such as trigger sensitivity and localization accuracy. Therefore, pre-launch simulation of the in-orbit background spectrum and count rate is essential for a realistic performance assessment of the detector.

The primary radiation environment for the BSD instrument on the CSS, in its low-Earth orbit (altitude $\sim$400 km, inclination $\sim$40$^\circ$), consists of several components. The major contributor is the cosmic X-ray background (CXB). Additional sources include cosmic rays (primarily protons, electrons, and positrons, along with secondary particles generated by their interactions with the atmosphere and spacecraft). Albedo gamma-rays from the Earth's atmosphere are excluded from this study, as the BSD's field of view is oriented towards space, and these particles are largely blocked by the station structure. When passing through the SAA region, the payload would be powered off. However, the high-flux of high-energy protons and electrons can activate materials in the payload or the CSS. The de-excitation of these materials produces considerable radiation and some characteristic emission lines. After exiting the SAA, when the payload is powered on, an amount of delayed background radiation may still be detected~\citep{he2020,liao2023}. This radiation is complex and orbit-dependent, and further detailed study of the in-orbit background will be conducted subsequently.

CXB, attributed to the integrated emission of extragalactic point sources, is modeled as an isotropic component with a broken power-law spectrum \citep{GEHRELS1992513}, given by Equation \ref{eq:cxb_bk}. The spectra for primary cosmic rays (protons, electrons, and positrons) are adopted from measurements by the AMS experiment \citep{200010, 2000215}, with empirical models implemented as described in \citep{Mizuno_2007}. Albedo neutrons, generated by cosmic ray interactions with the atmosphere, are also considered due to their potential for capture in the GAGG crystals. Their spectrum is derived from a segmented power-law fit to COMPTEL data \citep{https://doi.org/10.1029/JA078i016p02715}.

\begin{equation}
\label{eq:cxb_bk}
    \frac{dN_{\rm CXB}}{dE} = 
    \begin{cases}
        0.54E^{-1.4}    &,\;  E < 0.02 \text{ MeV} \\
        0.0117E^{-2.38} &,\;  0.02 \leq E < 0.1 \text{ MeV} \\
        0.014E^{-2.3}   &,\;  E \geq 0.1 \text{ MeV}
    \end{cases}
\end{equation}

\noindent where the $dN_{\rm CXB}/dE$ is expressed in units of counts~$\cdot$~cm$^{-2}~\cdot$~s$^{-1}~\cdot$~MeV$^{-1}~\cdot$~sr$^{-1}$.

\begin{figure}[h]
\centering
\includegraphics[width=0.55\textwidth]{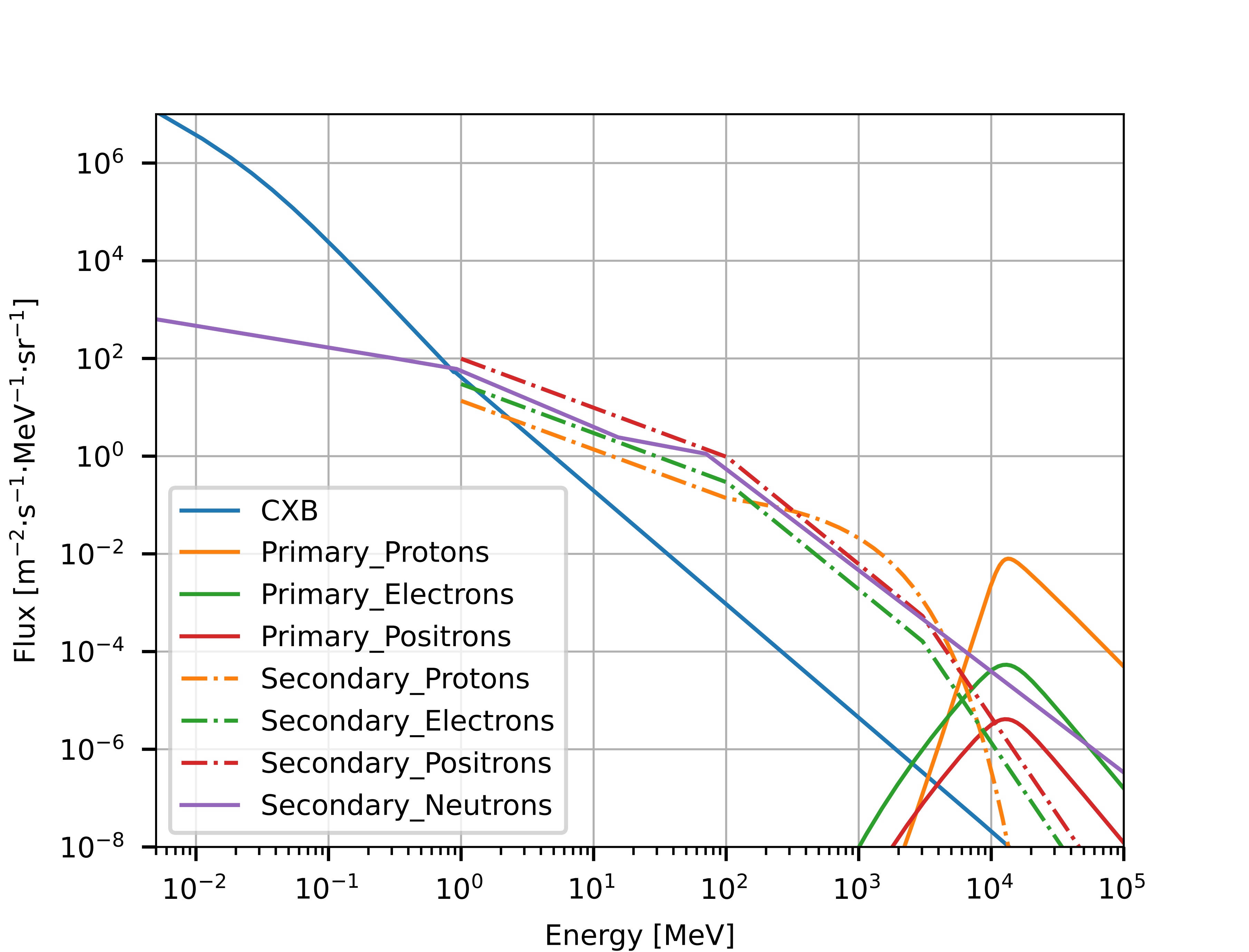}
\caption{Spectral models of the space radiation environment components considered for the BSD in-orbit background simulation} 
\label{fig:5}
\end{figure}

\begin{figure}[h]
\centering
\includegraphics[width=0.55\textwidth]{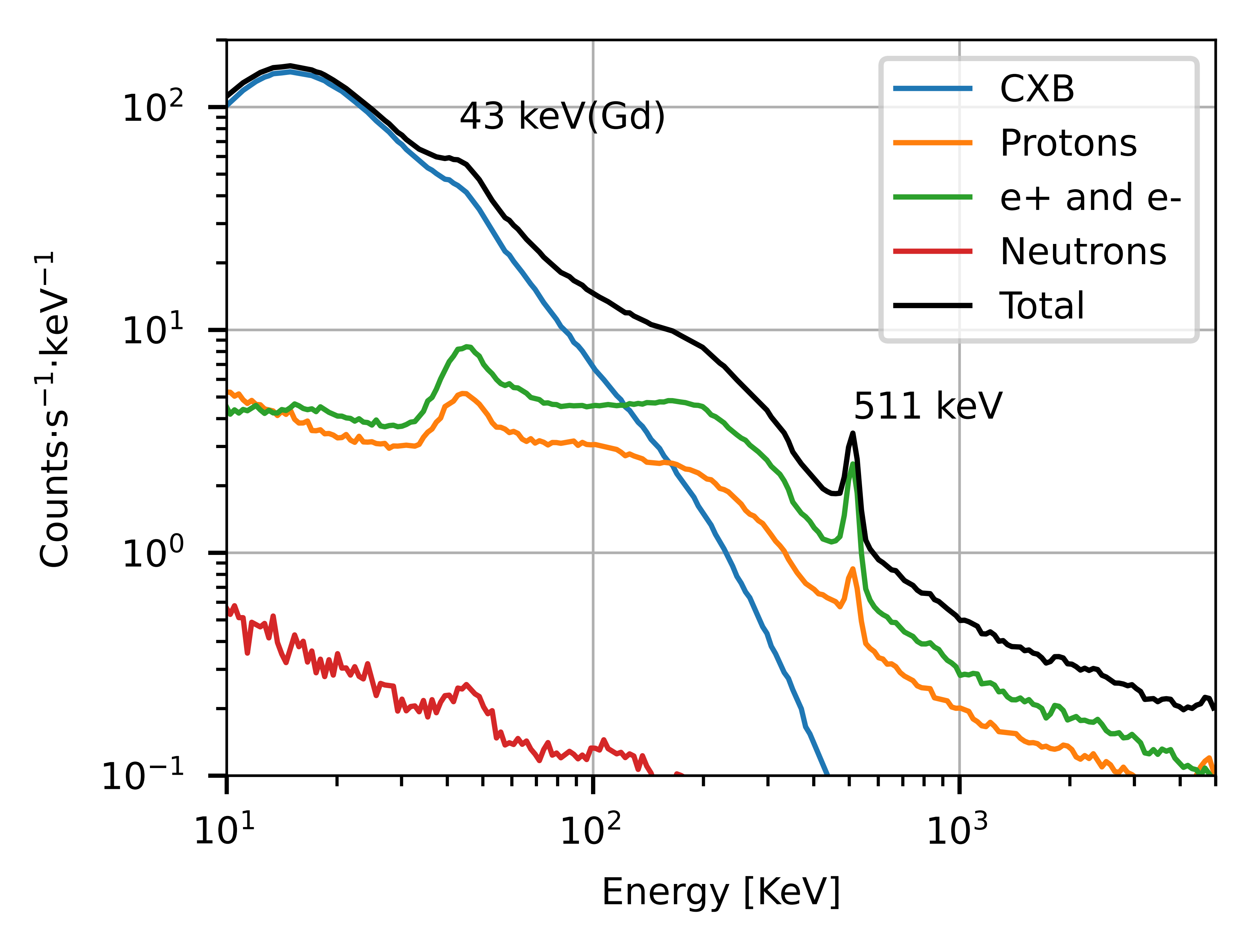}
\caption{Simulated energy deposition spectra in the BSD detector from individual in-orbit background components} 
\label{fig:6}
\end{figure}

The input spectra for these background components are shown in Fig.~\ref{fig:5}. The resulting simulated energy deposition spectra in the BSD detector are presented in Fig.~\ref{fig:6}. The corresponding total and segmented count rates for each component are summarized in Table~\ref{tbl:2}.

\begin{table}[h]
\caption{Simulated in-orbit background count rates for the BSD instrument}\label{tbl:2}
\begin{tabular}{@{}llll@{}}
\toprule
\multirow{2}{*}{Components} & \multicolumn{2}{c}{Count rate (counts/s)}\\
& 10-100~keV & 10-2000~keV\\
\midrule
CXB & 3942 & 4409 \\
Primary Proton & 291 & 1088 \\
Primary Electron & 7 & 28 \\
Primary Positron & 1 & 3 \\
Secondary Proton & 51 & 194 \\
Secondary Electron & 87 & 395 \\
Secondary Positron & 366 & 1658 \\
Neutron & 18 & 49 \\
Total & 4763 & 7824 \\
\bottomrule
\end{tabular}
\end{table}

Analysis of Table~\ref{tbl:2} indicates that CXB is the dominant background component below $\sim$150 keV. At higher energies, the background is primarily dominated by secondary particles. In contrast, the direct contributions from primary cosmic-ray electrons and positrons are negligible compared to other components.

These simulated in-orbit background results establish a critical foundation for subsequent, detailed investigations into the BSD instrument's performance capabilities within a realistic flight environment.

\subsection{ARF and RMF}\label{subsec:4-3}

The energy response matrix of a detector characterizes the relationship between the energy of incident photons and the resulting distribution of deposited energy within the detector. In X-ray astronomy, this matrix is conventionally separated into two components: the Redistribution Matrix File (RMF) and the Ancillary Response File (ARF). The RMF is a two-dimensional matrix that describes the probability for a photon of a given incident energy to be recorded in a specific energy channel. The ARF is a one-dimensional function that encapsulates the effective area of the detector as a function of the incident photon energy. Both the RMF and ARF are typically dependent on the incident angle of the photons.

\begin{figure}[h]
\centering
\includegraphics[width=0.65\textwidth]{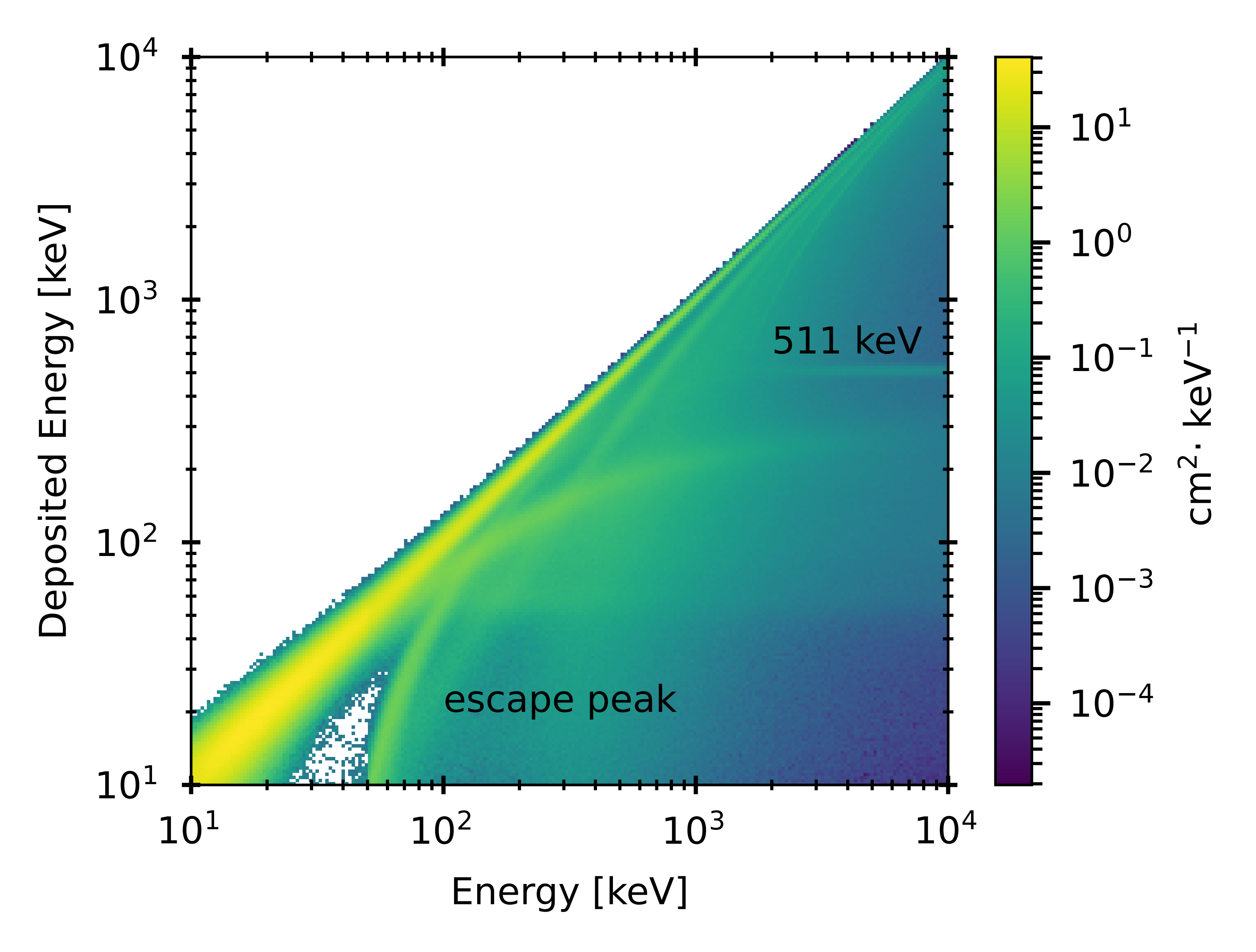}
\caption{RMF of BSD for on-axis incidence. The full-energy peaks, Gadolinium escape peaks, and the 511 keV electron-positron annihilation line are indicated} 
\label{fig:7}
\end{figure}

\begin{figure}[h]
\centering
\includegraphics[width=0.55\textwidth]{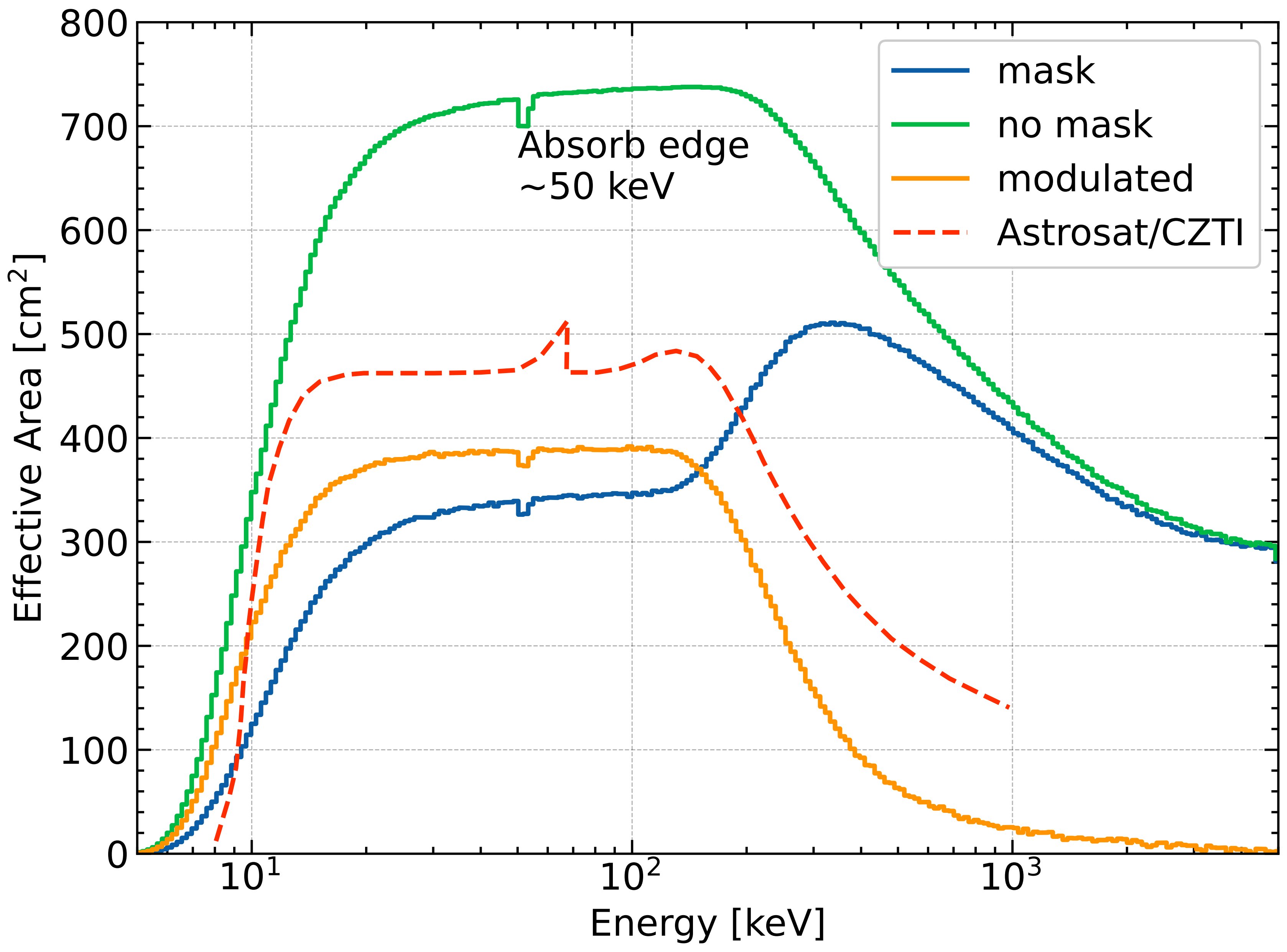}
\caption{Effective area of the BSD detector for on-axis incidence. The blue, green, and yellow curves correspond to the configurations with the coded-aperture mask, without the mask, and their difference (net modulated area), respectively. The difference quantifies the modulation and absorption introduced by the mask. The red dashed line represents the effective area of Astrosat/CZTI, while BSD shows greater sensitivity at higher energies} 
\label{fig:8}
\end{figure}

The energy response of the BSD CAMI detectors was investigated through Monte-Carlo simulations using the Geant4 mass model detailed in Section~\ref{subsec:4-1}. Simulations were performed for photons with various incident energies. Figure~\ref{fig:7} shows the simulated RMF for incoming on-axis photons in the energy range of 10 keV to 10 MeV. Figure~\ref{fig:8} compares the on-axis ARFs for the BSD instrument with and without the coded-aperture mask, along with their difference curve, over the energy range of 5~keV to 5~MeV. As a comparison, the effective area of AstroSat/CZTI\footnote{\url{https://astrosat.iucaa.in/czti/home}} which is also a CAMI instrument, is added in the Fig.~\ref{fig:8}. Compared to AstroSat/CZTI, BSD has a larger effective area above $\sim$200 keV.

Analysis of Fig.~\ref{fig:8} reveals that the effective area of BSD decreases at lower energies, primarily due to absorption by the coded-aperture mask plate. The maximum total effective area is approximately 500~cm$^{2}$. The effective area also decreases at higher energies, a consequence of the increasing transparency of the GAGG:Ce scintillator crystals to high-energy photons.

\subsection{Sensitivity and GRB detection rate}\label{subsec:4-4}

The instrument sensitivity and expected GRB detection rate were evaluated using the in-orbit background simulations (Section~\ref{subsec:4-2}) and the energy response matrices (Section~\ref{subsec:4-3}). The sensitivity is defined as the minimum source flux required to achieve a detection significance of $5\sigma$, and the significance $\text{Sig}$ is calculated using the simplified formula of Equation~\ref{eq:sig}.

\begin{equation}
\text{Sig} = \frac{S}{\sqrt{B}},
\label{eq:sig}
\end{equation}

\noindent where $B$ and $S$ represent the net background and source counts within the same observation time interval and energy range, respectively.

\begin{figure}[h]
\centering
\includegraphics[width=0.5\textwidth]{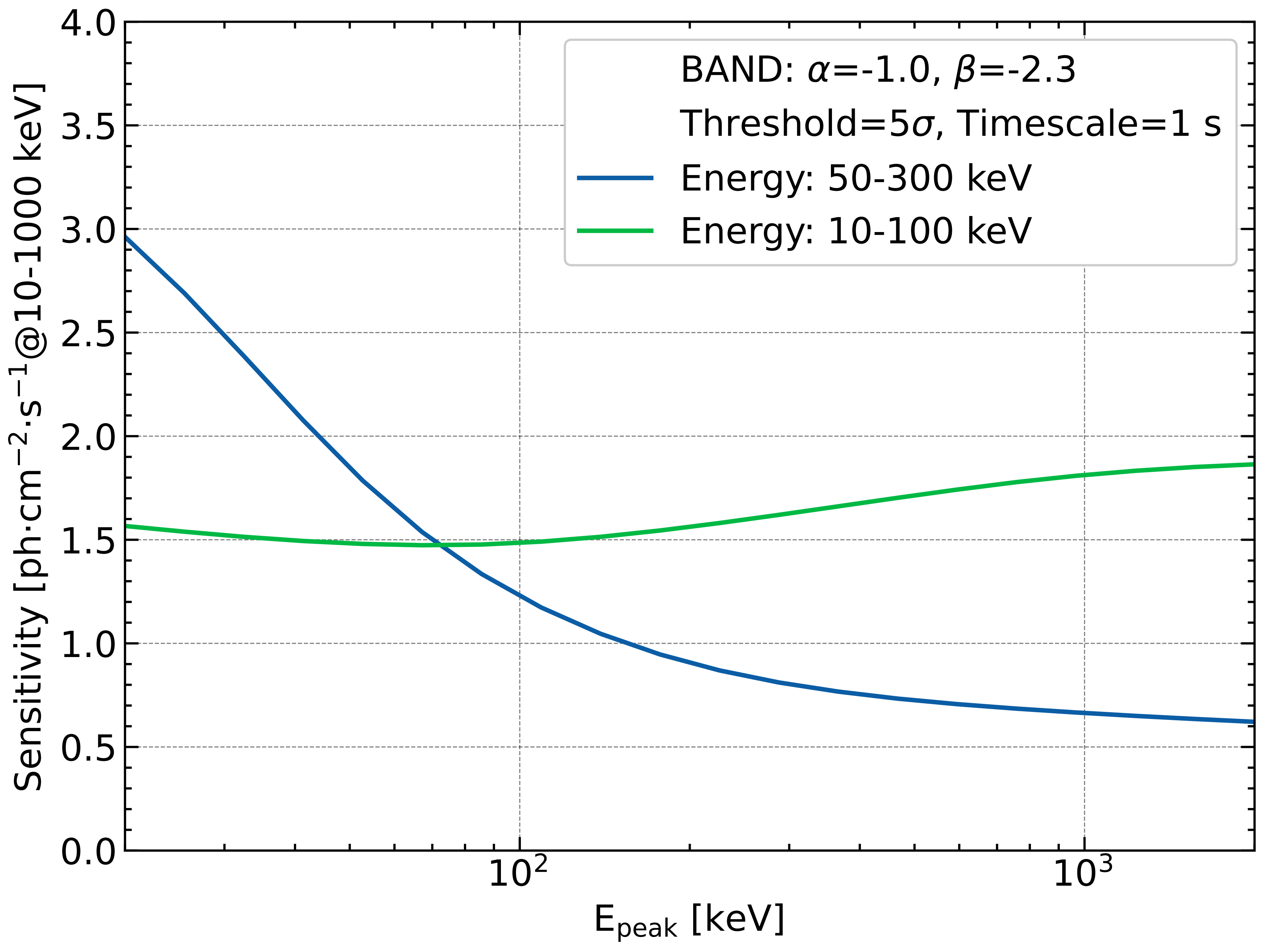}
\caption{Sensitivity of the BSD instrument for on-axis GRBs as a function of \(E_\mathrm{peak}\), evaluated in the 10--100 keV and 50--300 keV energy bands}
\label{fig:9}
\end{figure}

Using the Band function~\citep{band1993ApJBatse} with fixed spectral parameters ($\alpha = -1.0$, $\beta = -2.3$), the sensitivity of BSD was computed for different values of $E_\mathrm{peak}$, as shown in Fig.~\ref{fig:9}. The analysis considered two energy bands: the standard 10--100 keV range typical for GRB prompt emission, and an optimized 50--300 keV band chosen to enhance the sensitivity.

To provide a meaningful reference for evaluating the performance of BSD in GRB spectroscopy, we compare its sensitivity with that of AstroSat/CZTI, which has reported some GRB polarization measurements and operates on a similar Compton scattering detection principle. This comparative analysis reveals distinct behaviors between the two instruments. The sensitivity of CZTI, calculated based on its effective area (Fig.~\ref{fig:8}) and documented trigger threshold \citep{2021Sharma,2021Mate}, remains relatively stable across a range of $E_\mathrm{peak}$ values, maintaining approximately 0.66~ph~$\cdot$~cm$^{-2}$~$\cdot$~s$^{-1}$ at $E_\mathrm{peak} \sim 230$~keV. In contrast, the sensitivity of BSD not only achieves a competitive value of $\sim$0.8~ph~$\cdot$~cm$^{-2}$~$\cdot$~s$^{-1}$ in the 50--300 keV band at the same $E_\mathrm{peak}$, but also exhibits a marked improvement with increasing $E_\mathrm{peak}$. This highlights the enhanced performance of BSD for high-energy GRB spectroscopy and its complementary observational role.

\begin{figure}[h]
\centering
\includegraphics[width=0.54\textwidth]{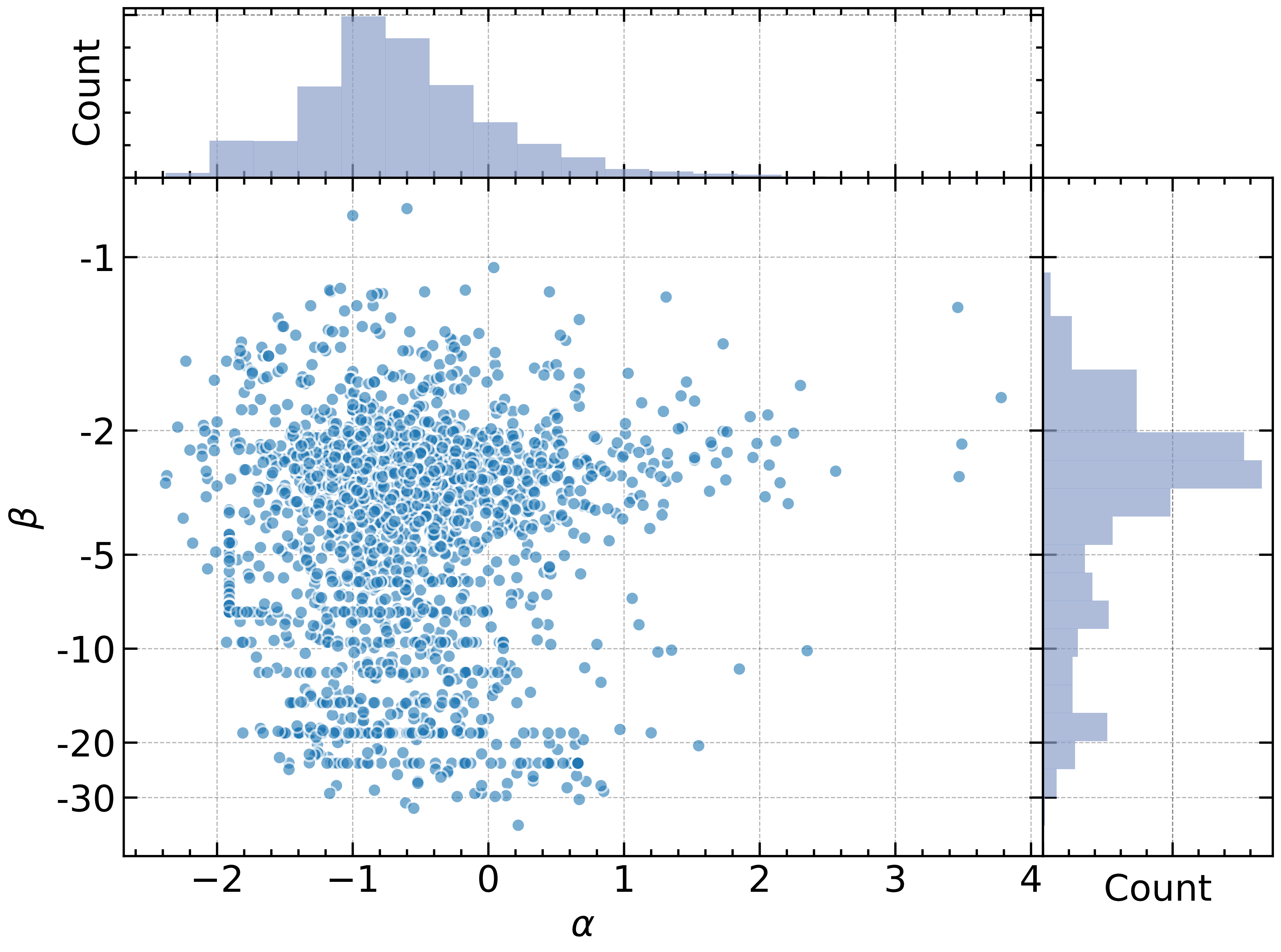}
\caption{Distribution of the Band function spectral parameters $\alpha$ and $\beta$ obtained from peak flux analysis of GRBs in the BATSE catalog} 
\label{fig:10}
\end{figure}

\begin{figure}[h]
\centering
\includegraphics[width=0.53\textwidth]{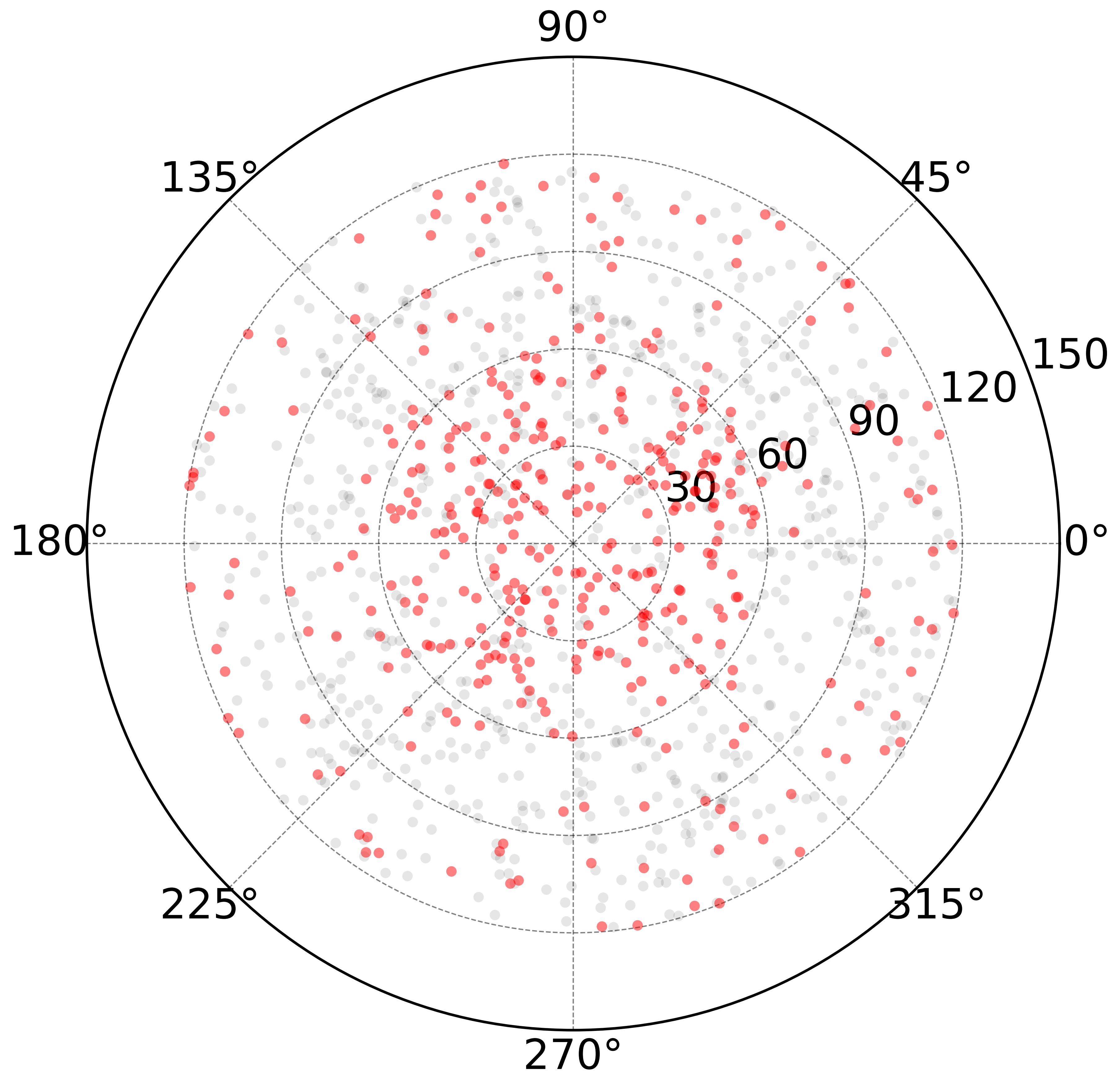}
\caption{Simulated distribution of GRBs in the BSD instrument coordinate system, with events exceeding the $5\sigma$ detection threshold highlighted in red} 
\label{fig:11}
\end{figure}

To estimate the expected GRB detection rate, we generated a simulated sample of 1650 GRBs, corresponding to three years of observation assuming an annual rate of 550 GRBs \citep{Band2002}. The spectral parameters for these simulated bursts were randomly drawn from the distributions of $\alpha$ and $\beta$ derived from the CGRO/BATSE catalog \citep{goldstein2013}, as shown in Fig.~\ref{fig:10}. For each simulated GRB, we determined its visibility from the CSS orbit, excluding periods when the GRBs are obstructed by the Earth or affected by the SAA region where high particle flux is considered. The resulting sky distribution of simulated GRBs relative to the BSD instrument is shown in Fig.~\ref{fig:11}, with red markers indicating the detectable events.

The simulation results show that BSD is expected to trigger on approximately 127 GRBs per year, with about 78 of these occurring within the $120^\circ$ partially-coded field of view (Fig.~\ref{fig:11}).

\subsection{GRB localization capability}\label{subsec:4-5}

According to the requirements for POLAR-2 polarization analysis, BSD should provide GRB localizations with an accuracy better than $1^\circ$ for typical bright GRBs. Distinct localization approaches have been employed by previously launched GRB detection instruments, including CGRO/BATSE, Swift/BAT, Fermi/GBM, GECAM, SVOM/GRM and SVOM/Eclairs, etc. Instruments such as BATSE, GBM, GECAM and GRM utilize the count-rate comparison method across multiple wide-field detectors oriented in different directions. While this technique provides large FoV coverage, its localization accuracy is typically limited to a few degrees at best~\citep{zhaoyi2023}.

In contrast, the CAMI method offers an optimal balance between wide FoV and high angular resolution, making it the preferred solution for the BSD instrument. This technique is capable of achieving arcminute-level precision while maintaining sensitivity across a broad energy range from hard X-rays to soft gamma-rays.

The fundamental principle of CAMI involves capturing different projections of the sky through a patterned mask. The recorded detector counts $\mathbf{D}$ can be expressed as:

\begin{equation}
\mathbf{D} = \mathbf{S} \ast \mathbf{M} + \mathbf{B},
\end{equation}

\noindent where $\mathbf{S}$ represents the sky image, $\mathbf{M}$ is the mask pattern, and $\mathbf{B}$ denotes the background counts. Image reconstruction is performed using a balanced correlation decoding algorithm. The decoded sky image is obtained through:

\begin{equation}
\label{eq:decode}
S_{ij} = \frac{\sum_{kl}{G^{+}_{i+k,j+l}W_{kl}D_{kl}}}{\sum_{kl}{G^{+}_{i+k,j+l}W_{kl}}} - \frac{\sum_{kl}{G^{-}_{i+k,j+l}W_{kl}D_{kl}}}{\sum_{kl}{G^{-}_{i+k,j+l}W_{kl}}},
\end{equation}

\noindent where $\mathbf{G}^+$ and $\mathbf{G}^-$ represent the balanced correlation mask and anti-mask patterns, respectively, and $\mathbf{W}$ contains the detector weighting coefficients \citep{goldwurm+03}. The source position is then determined using a centroid-fitting algorithm applied to the reconstructed image.

\begin{figure}[h]
\centering
\includegraphics[width=0.9\textwidth]{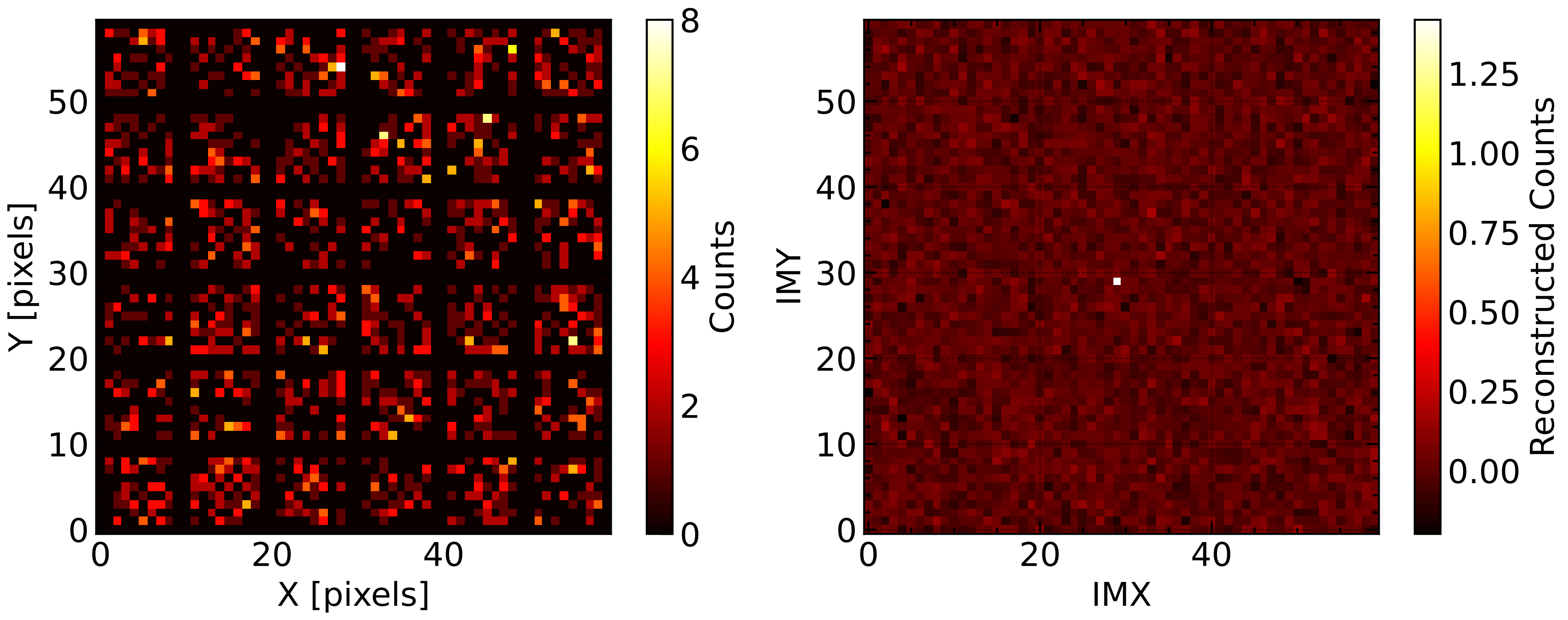}
\caption{Demonstration of the coded-aperture mask imaging process. Left: simulated detector count pattern from a GRB. Right: reconstructed image where the white point shows the source position in the imaging coordinate} 
\label{fig:12}
\end{figure}

To evaluate the localization performance of the BSD design, we conducted comprehensive Geant4 simulations of GRBs with varying fluences. Figure~\ref{fig:12} shows a representative example for the on-axis case, displaying both the recorded detector count pattern and the reconstructed image for a GRB with a fluence of $1\times10^{-6}~erg~\cdot~cm^{-2}$ (integrated over 10–1000~keV) in 1~s. The clean reconstruction demonstrates the effectiveness of the balanced correlation decoding algorithm in resolving source positions.

\begin{figure}[h]
\centering
\includegraphics[width=0.7\textwidth]{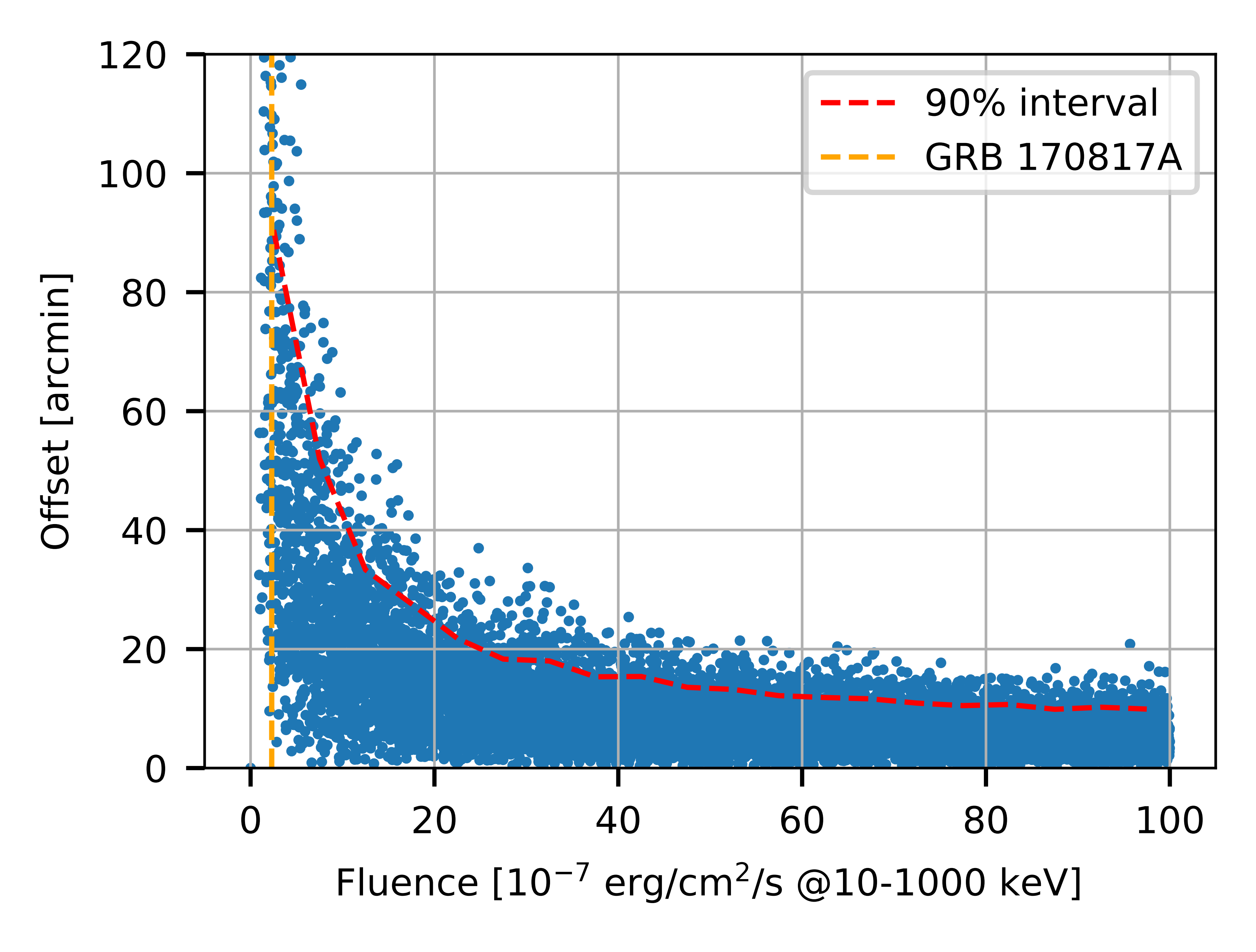}
\caption{Localization accuracy as a function of GRB fluence. The distribution of angular offsets is shown for multiple realizations at different fluence levels, with the red dashed line indicating the 90\% confidence region, the yellow dashed line indicates the position of GRB~170817A~\citep{2017GCNKienlin}. The input GRB spectrum is modeled with a Band function ($\alpha = -1.0$, $\beta = -2.3$, $E_\mathrm{peak} = 230$~keV) }
\label{fig:13}
\end{figure}

The systematic evaluation of localization accuracy as a function of fluence is presented in Fig.~\ref{fig:13}. At the lowest fluences considered ($\sim 10^{-7}~erg~\cdot~cm^{-2}$), the localization accuracy approaches $\sim2^{\circ}$, while for brighter bursts with fluences above $10^{-6}~erg~\cdot~cm^{-2}$, the offsets between the reconstructed and input source positions are kept to below $1^{\circ}$. For faint GRBs comparable to GRB~170817A in flux, the simulated localization accuracy is approximately $1.5^\circ$, as shown in Fig.~\ref{fig:13}. The demonstrated localization capability will allow POLAR-2 to independently provide the accurate source positions required for reliable polarization measurements, significantly expanding the sample of GRBs with high-quality polarimetry data compared to previous missions. This robust performance validates the current mask-to-detector separation and pixel size optimization, confirming that BSD will serve as a reliable instrument for GRB localization in the POLAR-2 mission.

\subsection{Polarimetry with BSD}\label{subsec:4-6}

\begin{figure}[h]
\centering
\includegraphics[width=0.55\textwidth]{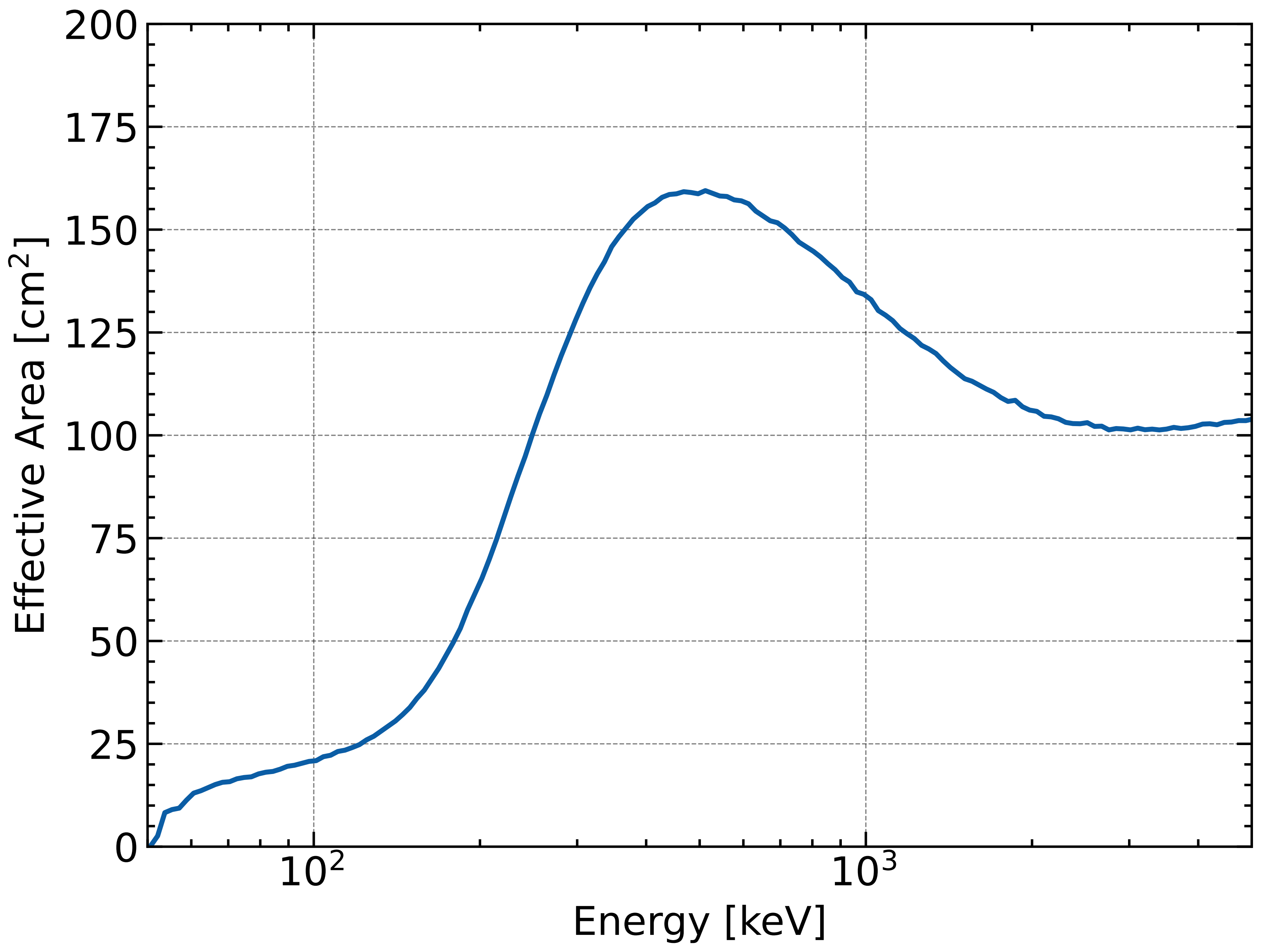}
\caption{Polarimetry effective area of the BSD instrument for Compton double-hit events, showing a peak sensitivity around 450 keV}
\label{fig:14}
\end{figure}

In addition to its capabilities in GRB spectroscopy and high-precision localization, BSD naturally possesses inherent polarimetric sensitivity due to its large-area pixelated detector array design—a philosophy shared by instruments like AstroSat/CZTI, Daksha~\citep{2023Bala}, etc. Our analysis indicates that the use of GAGG, a high-Z scintillator material, enables polarization measurements at higher energy ranges compared to LPD and HPD. High-energy gamma-ray photons undergoing Compton scattering in one GAGG crystal pixel are likely to be subsequently absorbed by other pixels in the array. This characteristic provides BSD with significant potential for simultaneous measurement of both polarization and energy of gamma-ray photons, making it promising for studying energy-dependent polarization evolution in GRBs.\\
\indent The polarimetric capability of BSD will serve as an important complement to LPD and HPD, warranting further in-depth exploration. To quantify this capability, we first computed the effective area for polarization measurements using the BSD detector array. This was done by selecting events where incident gamma-ray photons undergo at least two interactions within the GAGG crystal array, with at least one of these interactions being a Compton scattering process. Preliminary Monte-Carlo simulation results for the polarimetric effective area of the BSD detector, assuming an incident photon angle of \(0^\circ\), are shown in Fig.~\ref{fig:14}. The results demonstrate that the polarimetric effective area of BSD reaches its maximum at approximately 450 keV. Compared to the primary energy range of 100–300 keV for polarization measurements in CZTI~\citep{2014chattopadhyay}, BSD exhibits higher sensitivity to GRB polarimetry of higher-energy photons.

\begin{figure}[h]
\centering
\includegraphics[width=0.55\textwidth]{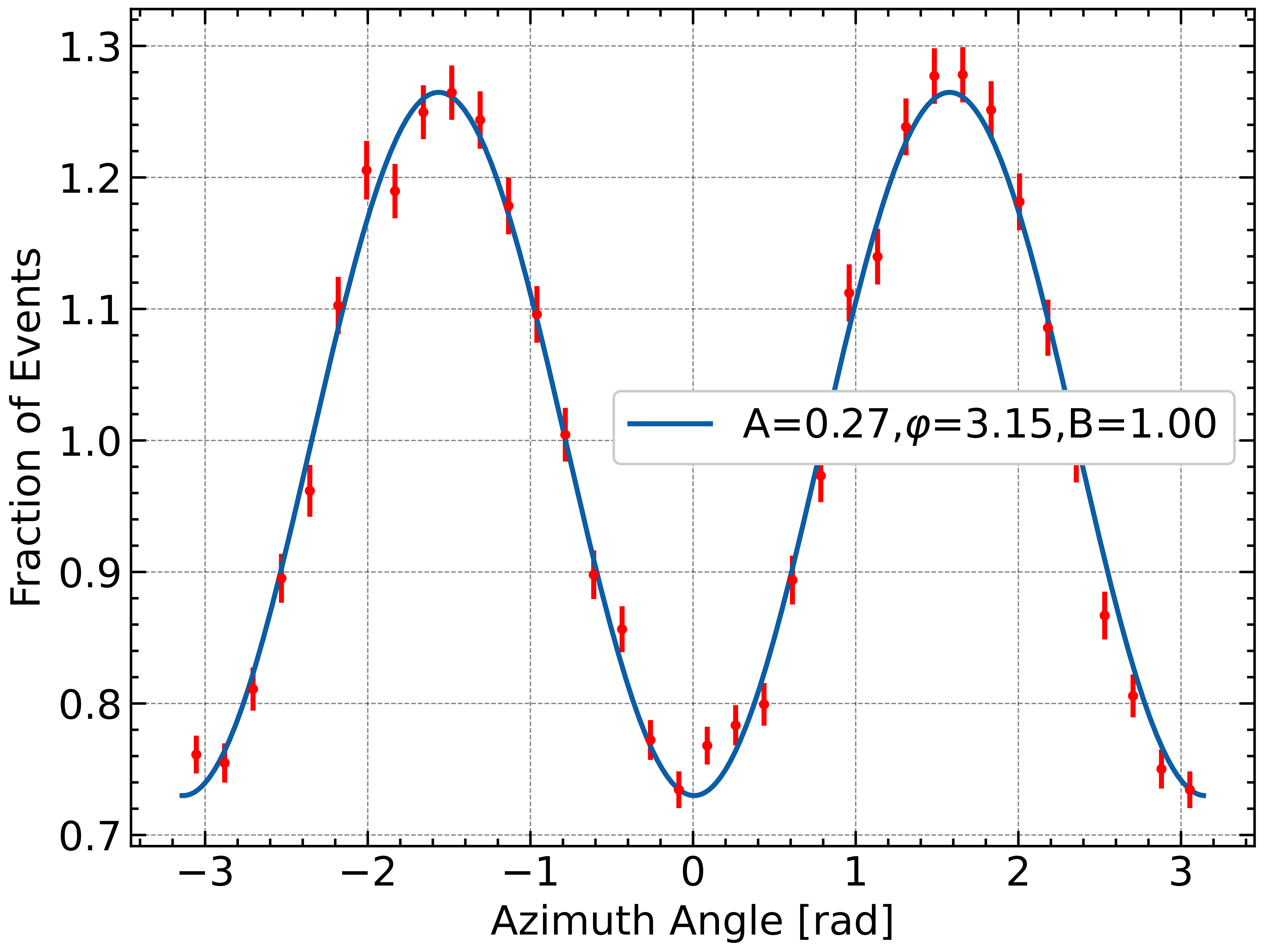}
\caption{Simulated modulation curve at 450~keV, normalized with the unpolarized data to remove instrumental effects}
\label{fig:15}
\end{figure}

The simulated azimuthal scattering angle distribution for 450~keV photons with the $0^{\circ}$ incident angle is presented in Fig.~\ref{fig:15}. The distribution is normalized to that expected for an unpolarized source. A modulation of the form given in Equation~\ref{eq:modulation_curve} is clearly observed. From the fit, a modulation factor of $\mu_{100} = A/B = 0.27$ is derived, which characterizes the instrument's response to fully polarized radiation at this energy.

\begin{equation}
C(\eta) = A \cos(2(\eta - \varphi + \pi/2)) + B
\label{eq:modulation_curve}
\end{equation}

\begin{figure}[h]
\centering
\includegraphics[width=0.55\textwidth]{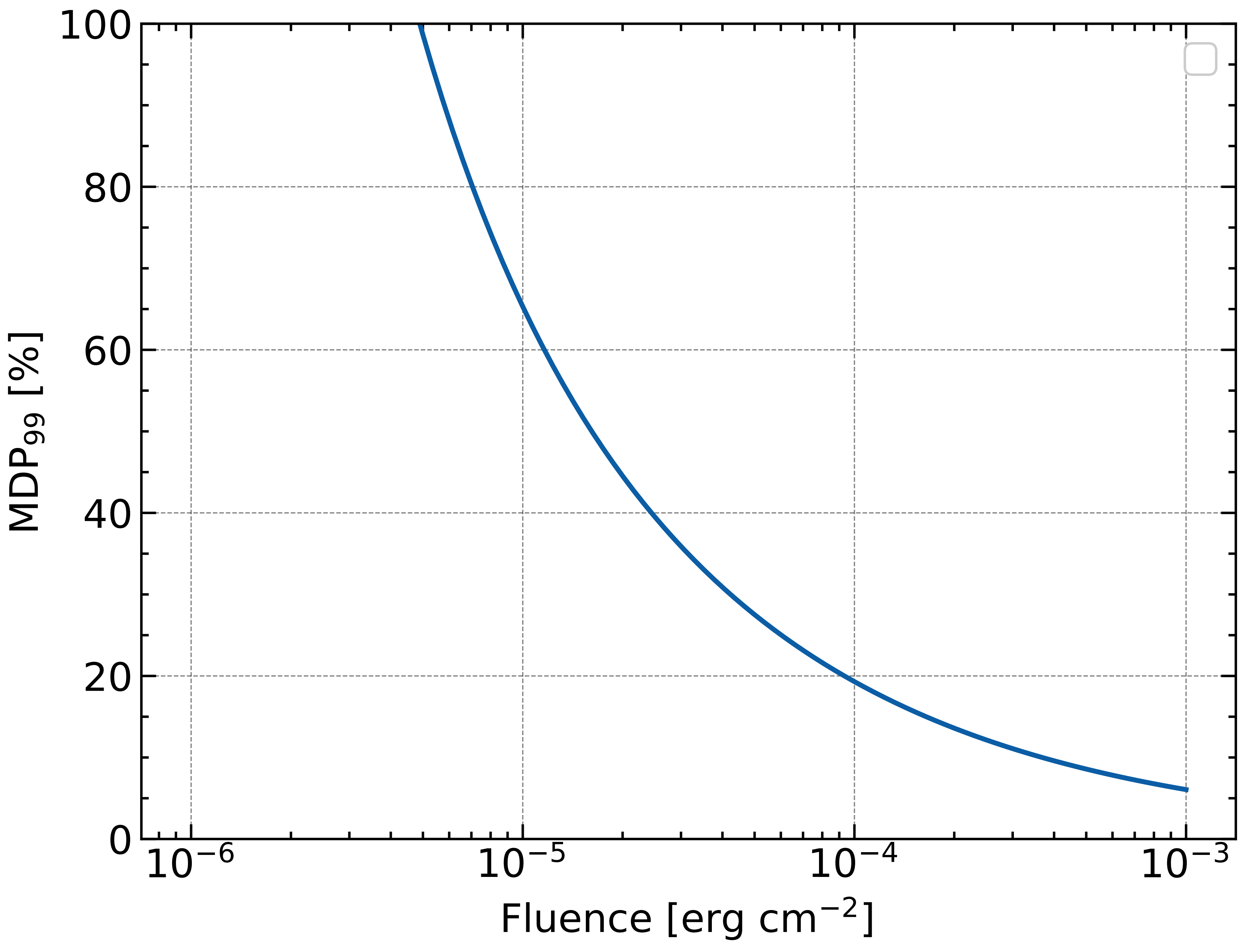}
\caption{The MDP as a function of fluence, calculated using the simulated in-orbit background (Fig.~\ref{fig:6}) and the Compton effective area (Fig.~\ref{fig:14}). The MDP at the 50$\%$ corresponds to $\sim 1.6 \times 10^{-5}$~erg~$\cdot$~cm$^{-2}$ in 1~s}
\label{fig:mdp}
\end{figure}

The Minimum Detectable Polarization (MDP)~\citep{2010SPIE.7732E..0EW} was computed using Equation~\ref{eq:mdp}, incorporating the simulated modulation factor $\mu_{100}$, the polarimetry effective area (Fig.~\ref{fig:14}), and the in-orbit background estimated from the double-bar event rate (Section~\ref{subsec:4-2}). For on-axis incidence, a MDP of 50\% is achieved for a GRB with a fluence of $\sim 1.6 \times 10^{-5}$~erg~$\cdot$~cm$^{-2}$ (integrated over 10–1000~keV) and a duration of 1~s (Fig.~\ref{fig:mdp}).
\begin{equation}
MDP_{99\%} = \frac{4.29}{\mu_{100} R_{\text{src}}} \sqrt{\frac{R_{\text{src}} + R_{\text{bkg}}}{T}},
\label{eq:mdp}
\end{equation}

\section{Prototype development and tests}\label{sec:5}

\begin{figure}[h]
\centering
\includegraphics[width=0.6\textwidth]{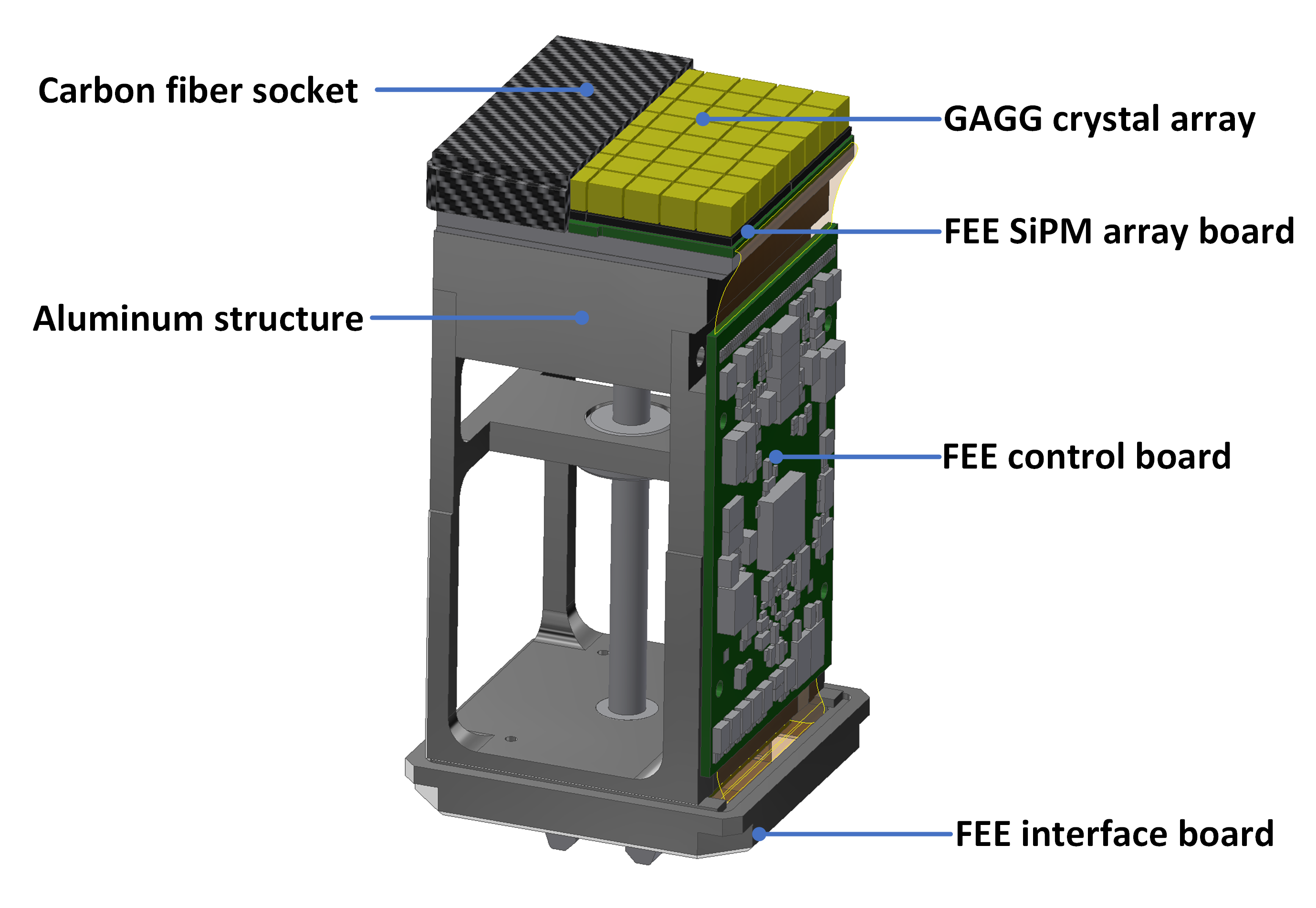}
\caption{Schematic design of the BSD detector module prototype for ESRF beam test} 
\label{fig:16}
\end{figure}

An $8 \times 8$ array of GAGG:Ce crystals, with individual pixel dimensions of $5.75 \times 5.75 \times 5$~mm, was developed for performance verification. This crystal array was coupled with a front-end electronics unit from the HPD payload to form an engineering model of a single BSD detector modular unit. The physical assembly of the prototype, which includes the crystal array, FEE, carbon fiber socket, and aluminum support structure, is shown in Fig.~\ref{fig:16}.

\begin{figure}[h]
\centering
\includegraphics[width=0.75\textwidth]{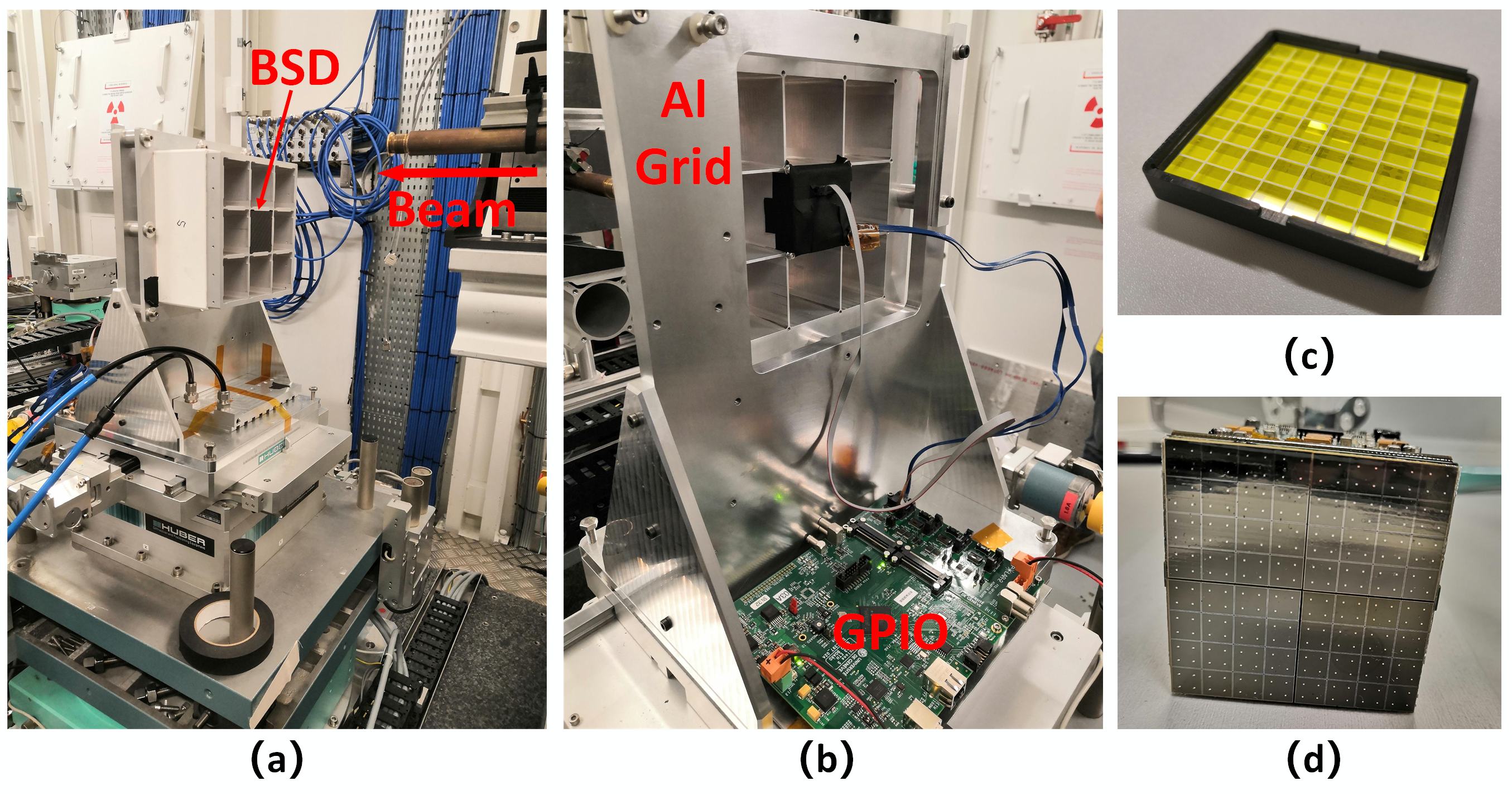}
\caption{Experimental setup for the BSD detector module prototype at the ESRF ID15A beamline. (a) Overview of the test configuration. (b) Rear view. (c) Close-up of the GAGG:Ce crystal array. (d) SiPM readout electronics board} 
\label{fig:17}
\end{figure}

The detector prototype was characterized using a hard X-ray beam at the ID15A beamline of the European Synchrotron Radiation Facility (ESRF) \citep{Vaughan:il5045}. A key objective was to verify that the GAGG-SiPM configuration achieves the 10 keV low-energy threshold—a critical requirement for capturing the low-energy X-ray component of GRB emission for accurately deriving physical parameters from more complete spectra and ensuring high photon statistics—as previously demonstrated with a GAGG-PMT system. The experimental setup, consistent with that used for polarization response tests of the HPD module \citep{Kole_2024}, is shown in Fig.~\ref{fig:17}. The prototype was mounted in an aluminum frame on a translation stage, enabling a horizontal and vertical scan across all 64 channels. During testing, a General Purpose Input/Output (GPIO) board was used for signal readout and front-end electronics (FEE) configuration, serving as a temporary interface pending the completion of the dedicated back-end electronics.

\begin{figure}[h]
\centering
\includegraphics[width=0.6\textwidth]{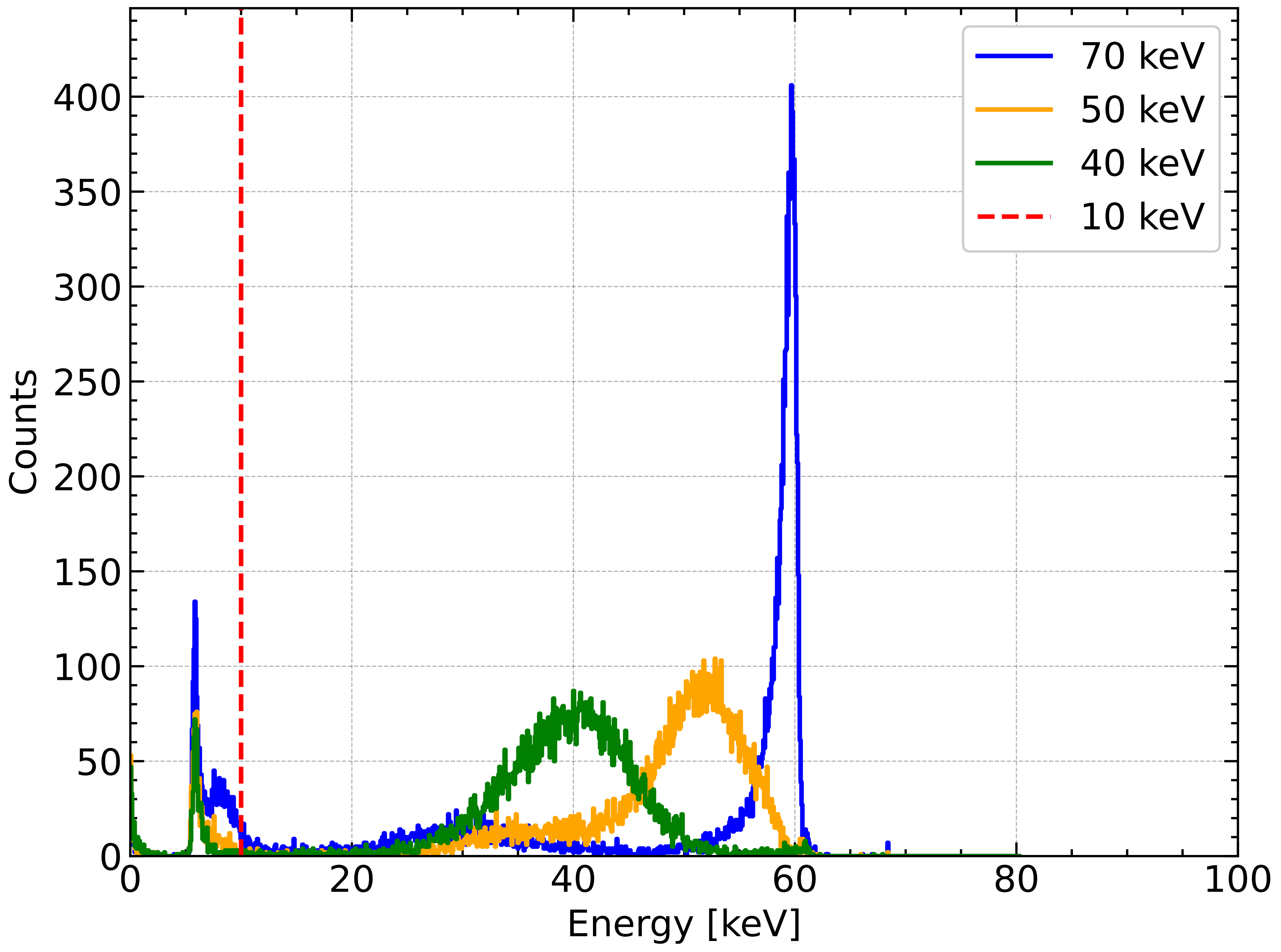}
\caption{Energy spectra of the BSD high-gain channel at 40, 50, and 70 keV beam energies} 
\label{fig:18}
\end{figure}

\begin{figure}[h]
\centering
\includegraphics[width=0.6\textwidth]{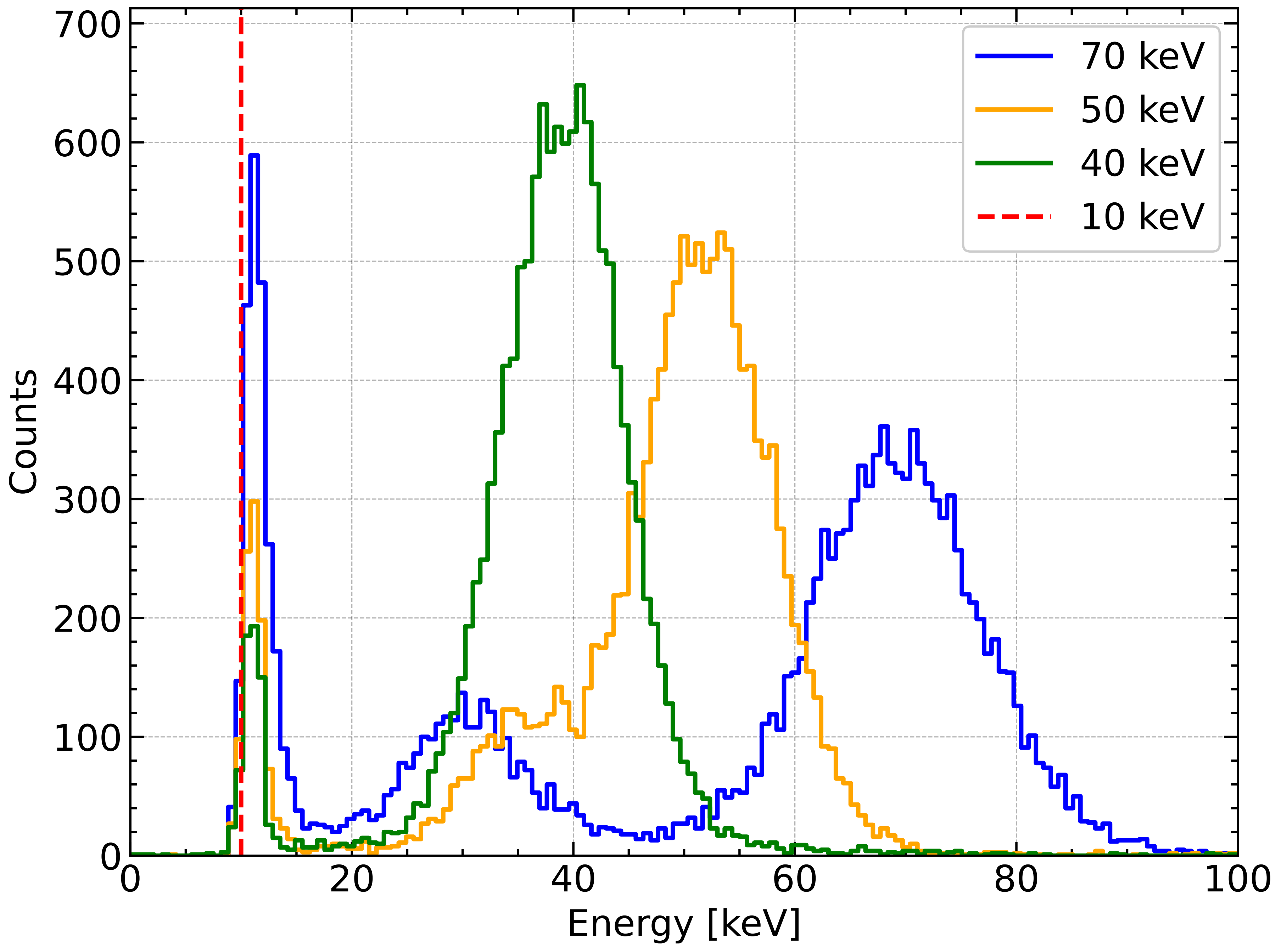}
\caption{Energy spectra of the BSD low-gain channel at 40, 50, and 70 keV beam energies} 
\label{fig:19}
\end{figure}

The energy response of this detector module prototype was evaluated at beam energies of 40, 50, and 70~keV. The results from these measurements are presented in Table.~\ref{tbl:3}. The energy spectra for the high-gain and low-gain channels after ADC-to-energy conversion are shown in Fig.~\ref{fig:18} and Fig.~\ref{fig:19}, respectively, with the red dashed line indicating the 10~keV reference. A spectral feature around 32 keV is visible in the spectra, attributed to barium (Ba) K$_\alpha$ fluorescence from the BaSO$_4$ coating in the prototype's GAGG crystal array housing. The mechanical housing material has been changed to Enhanced Specular Reflector (ESR) in the final flight design to prevent the X-ray photons to be absorbed in the dead volumes in between the GAGG crystals.

\begin{table}[h!]
\centering
\caption{The pulse height analysis (PHA) values (in units of ADC channel) and their resolution for the BSD high-gain and low-gain channels}
\label{tbl:3}
\begin{tabular*}{\textwidth}{l @{\extracolsep{\fill}} cccc}
\hline
\multirow{2}{*}{\textbf{Energy (keV)}} & \multicolumn{2}{c}{\textbf{High gain channel}} &\multicolumn{2}{c}{\textbf{Low gain channel}}\\
&PHA&Resolution(\%)&PHA&Resolution(\%)\\
\hline
40 & 2041&28.3 & 272&27.7\\
50 & -&- &349&24.3\\
70 & -&- &453&24.0\\
\hline
\end{tabular*}
\footnotetext[*]{For the high-gain channel, PHA values and resolution data for 50 and 70 keV are excluded due to the electronics saturation effects which was not optimized during the beam tests.}
\end{table}

The tests successfully demonstrated a low-energy threshold of 10~keV, achieved even with the SiPM operating at room temperature. This threshold is expected to improve during flight operations, as the planned operating temperature below $0\,^\circ\mathrm{C}$ will reduce SiPM dark noise. It should be noted that the detector parameters during this initial test were not fully optimized. For instance, a high gain setting caused signal saturation in some channels, limiting the effective dynamic range. Future optimization of parameters, such as the SiPM bias voltage and ASIC gain settings, is expected to significantly enhance the detector's performance.

\begin{figure}[h]
\centering
\includegraphics[width=0.5\textwidth]{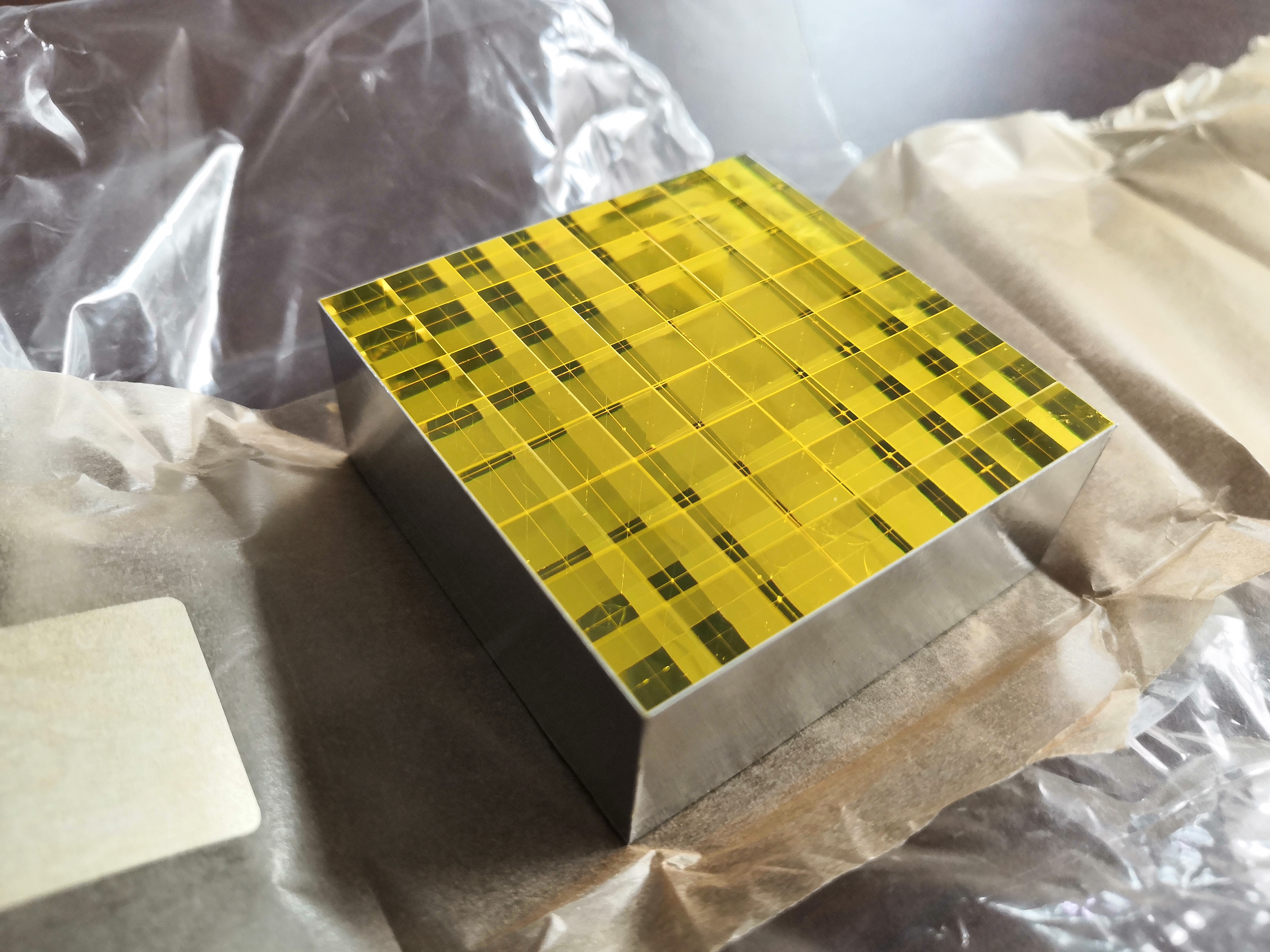}
\caption{Optimized GAGG:Ce crystal array with pixel dimensions of $5.75~mm \times 5.75~mm \times 20~mm$} 
\label{fig:20}
\end{figure}

\begin{figure}[h]
\centering
\includegraphics[width=0.5\textwidth]{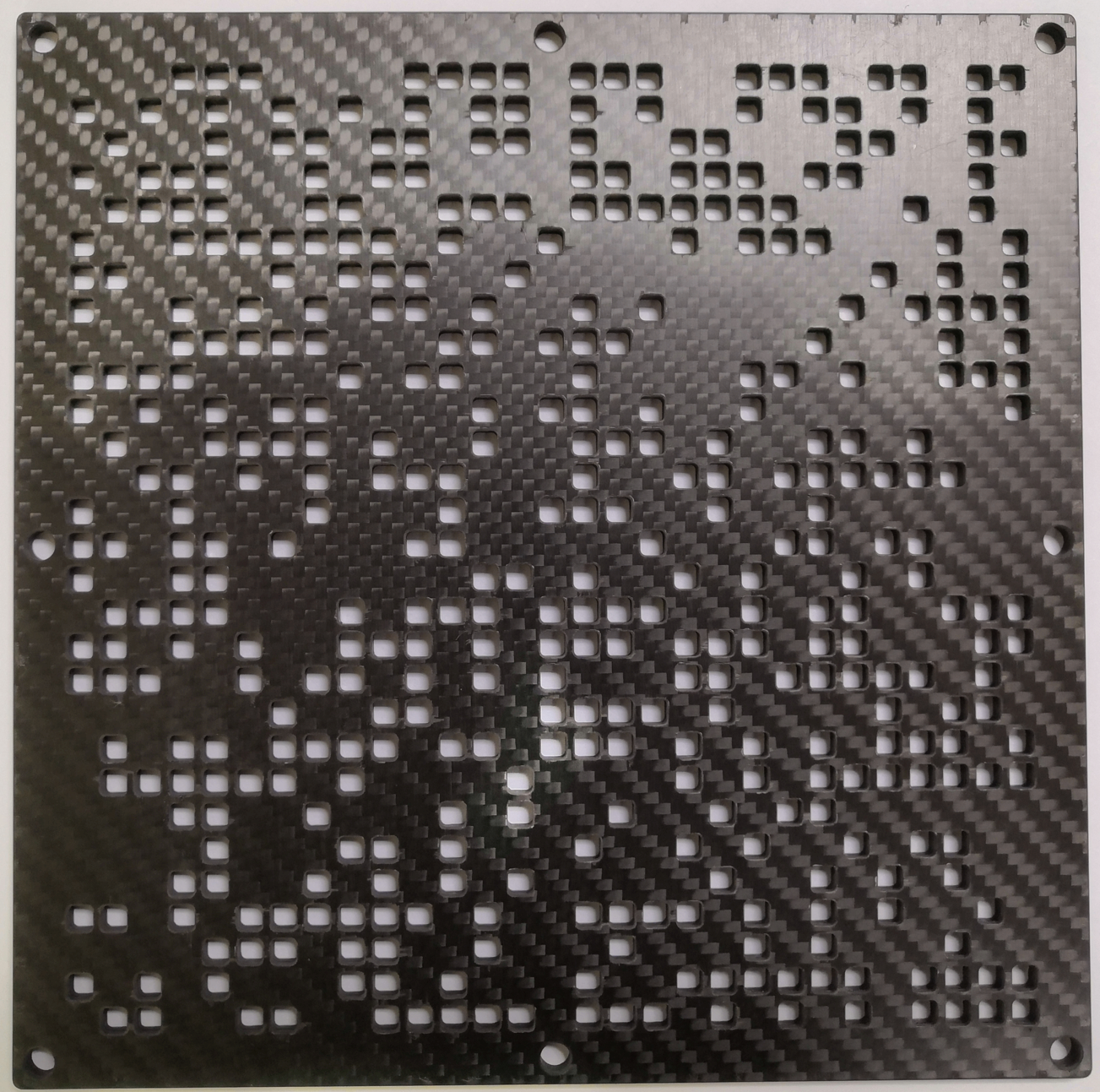}
\caption{A 1/9-scale model of the BSD coded-aperture mask plate, fabricated for design verification. The top carbon fiber layer with the aperture pattern (through which photons pass) is visible} 
\label{fig:21}
\end{figure}

Subsequent Monte-Carlo simulations led to an optimized crystal length of 20~mm (retaining the 5.75~mm$\times 5.75$~mm cross-section) to improve efficiency over the required energy range. A new crystal array with these dimensions has been fabricated (Fig.~\ref{fig:20}). Furthermore, a 1/9-scale coded-aperture mask plate has been produced (Fig.~\ref{fig:21}). Current and future work focuses on optimizing the detector parameters with the new crystal array and conducting performance tests.

\section{Summary and Conclusion}\label{sec:6}

The POLAR-2 mission, scheduled for installation on the China Space Station, now planned around 2028, is designed to conduct high-precision polarization, spectral, and temporal observations of high-energy transients, particularly GRBs. A key objective of POLAR-2 is to achieve significant breakthroughs in understanding the GRBs' radiation mechanisms, magnetic field configurations, and jet geometries. As an essential component of POLAR-2, BSD is tasked with providing precise source localization and broadband spectroscopy, which are critical for accurate polarization analysis.

This paper has detailed the design, simulated performance, and preliminary ground testing of the Broad-band Spectrometer Detector (BSD) instrument. The spectrometer adopts a coded-aperture mask imaging technique to meet its scientific requirements within the engineering constraints of the space station platform. A comprehensive performance evaluation was conducted using Geant4-based Monte-Carlo simulations, and the functionality of the detector concept was verified through prototype tests.

The simulation results confirm that the current design effectively fulfills the primary mission requirements. Specifically, BSD is capable of localizing typical GRBs with an accuracy better than $1^\circ$, enabling POLAR-2 to perform autonomous polarization analysis without complete reliance on external instruments. This will significantly increase the sample size of GRBs with high-quality polarization measurements. Furthermore, the design provides a large FoV to maximize the coincidence observation rate with the polarization detectors (HPD/LPD) and incorporates a real-time trigger and alert system to facilitate multi-messenger follow-up observations.

In terms of polarization measurement capability, our analysis demonstrates that the BSD instrument exhibits significant potential for detecting polarized signals from high-energy photons. While the CZTI's polarimetric sensitivity is optimized for the 100--300 keV energy range~\citep{2014chattopadhyay}, BSD shows a distinct advantage at higher energies due to its thicker GAGG:Ce crystals and larger pixel geometry. The calculated MDP$_{99\%}$ for BSD reaches 50\% at a fluence of $\sim 1.6 \times 10^{-5}$~erg~$\cdot$~cm$^{-2}$ in 1~s (Band:~$\alpha$ = -1.0, $\beta$ = -2.3, E$_\mathrm{peak}$ = 230~keV), which compares favorably with the performance of Daksha~\citep{2023Bala}. Although this sensitivity is somewhat inferior to the dedicated Compton telescope COSI, which achieves a 50\% MDP at $5 \times 10^{-6}$~erg~$\cdot$~cm$^{-2}$ in 1~s~\citep{2021tomsick}, BSD's polarimetry capability--when combined with its very large field of view (about 40\% of the sky in comparison with 25\% of the sky covered by COSI), spectroscopic and localization functions--creates a unique multi-messenger observational capacity. Besides, the HPD instrument of POLAR-2 achieves the same MDP at a fluence below $1 \times 10^{-6}$~erg~$\cdot$~cm$^{-2}$ in 1~s~\citep{Gill2021GRBPA}, demonstrating that the combined polarimetric sensitivity of the full POLAR-2 instruments represents a significant advancement in capability. Further optimization of the detector parameters and analysis techniques will be essential to fully exploit the scientific potential of BSD for gamma-ray polarimetry, particularly in its energy-resolved polarimetry capability.

Ground testing of an engineering prototype at the ESRF beamline demonstrated a low-energy threshold of 10 keV, meeting a critical requirement for broadband spectral measurements. However, these tests also revealed that the detector configuration parameters were not yet fully optimized.

In conclusion, the design and preliminary performance studies of the POLAR-2/BSD spectrometer confirm its feasibility and potential to become a high-performance instrument for GRB observations. The ongoing and planned optimization and testing activities will ensure that it achieves its scientific objectives upon deployment on the China Space Station.

\bmhead{Acknowledgements}

We acknowledge the European Synchrotron Radiation Facility (ESRF) for the provision of beam time and facilities essential to this work. We are particularly grateful to the ID15A beamline team for their invaluable assistance throughout the one-week experimental campaign. We acknowledge the 26th Research Institute of China Electronics Technology Group Corporation (Formerly Sichuan Institute of Piezoelectric and Acousto-optic Technologies, SIPAT) for provision of the GAGG crystal arrays. The front-end electronics were designed and developed by M.K. and N.D.A. using funding from the Swiss National Science Foundation through the Ambizione program.

\bmhead{Author Contributions} J-C. S., J. H., S-N. Z. and S-L. X. performed the analysis and simulations and were involved in paper writing and figures preparation. J-C. S., J. H., S-N. Z., S-L. X., J-T. L., Y-B. X., J. M., S. W., L. S., X-Z. L., H-B. L., F. X., M. Z., P. A., J. B., F. C., N. D. A., H-B. F., Z-K. F., M. G., J. G., A. G., J-X. H., Y. H., J. H., Z-H. J., M. K., D-W. L., H-C. L., T-L. L., L. P., A. P., N. P., D. R., A. S., L-M. S., C. T., X-M. W., Y-H. W., B-B. W., P-L. W., X. W., S. X., S. Y., L-Y. Z., L. Z. and Y-J. Z. contributed to the design, development and tests of the instrument. R. G. and J. G. contributed to the theoretical analysis part and instrument optimization. All authors reviewed the manuscript.

\bmhead{Funding} This work is supported by the Special Program for Enhancing Original Innovation Capability of Chinese Academy of Sciences (Grant No.~292024000260), Special Exchange Program A of Chinese Academy of Sciences (Grant No.~2H2025000112), the National Natural Science Foundation of China (Grant No.~11961141013), the Science and Technology Foundation of Guizhou Province (Key Program, Grant No.~[2025]021) and China's Space Origins Exploration Program.

\section*{Declarations}
\bmhead{Competing Interests} The authors declare no competing interests.

\bibliographystyle{plainnat}
\bibliography{sn-bibliography}

\nolinenumbers
\end{document}